\documentclass[fleqn,usenatbib]{mnras}

\usepackage{newtxtext,newtxmath}

\usepackage[T1]{fontenc}

\DeclareRobustCommand{\VAN}[3]{#2}
\let\VANthebibliography\thebibliography
\def\thebibliography{\DeclareRobustCommand{\VAN}[3]{##3}\VANthebibliography}

\usepackage{graphicx}	
\usepackage{amsmath}	
\usepackage{ulem}

\newcommand{\rev}[1]{\textcolor{black}{#1}}

\title[RGZ EMU: Semantic Morphology Taxonomy]{Radio Galaxy Zoo EMU: Towards a Semantic Radio Galaxy Morphology Taxonomy}

\author[M. Bowles et al.]{
Micah Bowles$^{1}$\thanks{E-mail: micah.bowles@postgrad.manchester.ac.uk},
Hongming Tang$^{2}$,
Eleni Vardoulaki$^{3}$,
Emma L. Alexander$^{1}$,
Yan Luo$^{4}$,
\newauthor
Lawrence Rudnick$^{5}$,
Mike Walmsley$^{1}$,
Fiona Porter$^{1}$,
Anna M.~M.~Scaife$^{1,6}$,
Inigo Val Slijepcevic$^{1}$,
\newauthor
Elizabeth A. K. Adams$^{7,8}$,
Alexander Drabent$^{3}$,
Thomas Dugdale$^{1}$,
G\"ulay G\"urkan$^{9,3,10}$,
\newauthor
Andrew M. Hopkins$^{11}$,
Eric F. Jimenez-Andrade$^{12}$,
Denis A. Leahy$^{13}$,
Ray P. Norris$^{14,15}$,
\newauthor
Syed Faisal ur Rahman$^{16}$,
Xichang Ouyang$^{4}$,
Gary Segal$^{17,15}$,
Stanislav S. Shabala$^{18}$,
O. Ivy Wong$^{10,19,20}$
\\
$^{1}$Jodrell Bank Centre for Astrophysics, Department of Physics and Astronomy, University of Manchester, Manchester, UK\\
$^{2}$Department of Astronomy, Tsinghua University, Beijing 100084, China\\
$^{3}$Th\"{u}ringer Landessternwarte, Sternwarte 5, 07778 Tautenburg, Germany\\	
$^{4}$School of Physics and Astronomy, Sun Yat-sen University, 2 Daxue Road, Zhuhai 519082, China\\
$^{5}$Minnesota Institute for Astrophysics, University of Minnesota, 116 Church St., SE, Minneapolis, MN 55455, USA\\
$^{6}$The Alan Turing Institute, Euston Road, London NW1 2DB, UK\\
$^{7}$ASTRON, the Netherlands Institute for Radio Astronomy, Oude Hoogeveesedijk 4, 7991 PD Dwingeloo, The Netherlands\\
$^{8}$Kapteyn Astronomical Institute, University of Groningen, PO Box 800, 9700 AV Groningen, The Netherlands\\
$^{9}$Centre for Astrophysics Research, University of Hertfordshire, Hatfield, AL10 9AB, UK\\
$^{10}$CSIRO Space and Astronomy, ATNF, PO Box 1130, Bentley WA 6102, Australia\\
$^{11}$Australian Astronomical Optics, Macquarie University, 105 Delhi Rd, North Ryde, NSW 2113, Australia\\
$^{12}$Instituto de Radioastronomía y Astrofísica, Universidad Nacional Autónoma de México, Antigua Carretera a Pátzcuaro \# 8701,\\ Ex-Hda. San José de la Huerta, Morelia, Michoacán, México C.P. 58089\\
$^{13}$Department of Physics and Astronomy, University of Calgary, Calgary, Canada\\
$^{14}$School of Science, Western Sydney University, Locked Bag 1797, Penrith, NSW 2751, Australia\\
$^{15}$CSIRO Space \& Astronomy, P.O. Box 76, Epping, NSW 1710, Australia\\
$^{16}$Institute of Space Science and Technology, University of Karachi\\
$^{17}$School of Mathematics and Physics, University of Queensland, St Lucia, Brisbane, QLD 4072, Australia\\
$^{18}$School of Natural Sciences, University of Tasmania, Private Bag 37, Hobart, TAS 7001, Australia.\\
$^{19}$ICRAR-M468, The University of Western Australia, 35 Stirling Hwy, Crawley, WA 6009, Australia\\
$^{20}$ARC Centre of Excellence for All-Sky Astrophysics in 3 Dimensions (ASTRO 3D), Australia\\
}

\date{Accepted XXX. Received YYY; in original form ZZZ}

\pubyear{2022}

\begin{document}
\label{firstpage}
\pagerange{\pageref{firstpage}--\pageref{lastpage}}
\maketitle

\begin{abstract}
We present a novel natural language processing (NLP) approach to deriving plain English descriptors for science cases otherwise restricted by obfuscating technical terminology. We address the limitations of common radio galaxy morphology classifications by applying this approach. We experimentally derive a set of semantic tags for the Radio Galaxy Zoo EMU (Evolutionary Map of the Universe) project and the wider astronomical community. We collect 8,486 plain English annotations of radio galaxy morphology, from which we derive a taxonomy of tags. The tags are plain English. The result is an extensible framework which is more flexible, more easily communicated, and more sensitive to rare feature combinations which are indescribable using the current framework of radio astronomy classifications.
\end{abstract}




\section{Introduction}

Language is often difficult to define or use. When new concepts arise and demand their own terminology, terms can be adopted from similar ideas \citep[e.\,g. `entropy' in information theory and physics;][]{Jordao2021Entropy}, invented \citep[e.\,g. `utopia';][]{Romm1991Utopia}, or named after the discoverers (e.\,g. `Newtonian physics'). Individual terms often have multiple accepted definitions within a given field, and especially across fields \citep[e.\,g. `modern' in philosophy and art;][]{Schelling1994modernPhilosophy,Thomas2022art-definition}. The construction of language in science is especially important, as language is believed to affect how we think \citep[][]{Wolff2010LinguisticRelativity}.

The terminology used in astronomy struggles with these same issues. 
Some terms are obfuscated, poorly defined, or so specific that experts can often only engage in the subject matter if a definition is provided on each use. In radio astronomy, a field which began in the 1930s \citep{Southworth1957RadioHistory}, the language used to describe celestial objects has been developed almost entirely in tandem with the instruments and corresponding scientific understanding. Consequently, some terms are limited by our physical understanding \citep[e.\,g. Little Green Man 1;][]{Hewish1968} or the sample inspected at the time \citep[e.\,g. FRI / FRII;][]{FR1974MNRAS.167P..31F}.

In the case of radio galaxy morphologies, language is becoming increasingly difficult to use, especially as technological and scientific advancements provide deeper insight into the vast range of radio morphologies. 
The gap between the diverse range of observed radio galaxy morphologies and the classification schemes used is widening. 
The current morphological classifications carry information which cannot be quantified under current frameworks, meaning the use of non-numeric features, i.\,e. language based schemes, is unavoidable.
Accordingly, the current classification schemes fall victim to obfuscated language. Additionally, the terms used tend to describe abstract classes, which lack the ability to capture the increasingly complex features of radio galaxies observed with the newest generation of instruments.
\citet{Rudnick+2021} urges the radio astronomy community to develop a tagging system rather than forcibly attempting to create classes which neatly separate objects. Such a tagging system would allow an object to be assigned plain English descriptors capturing the semantics of the object's features through tags rather than be assigned a distinct class to which it belongs. 
Additionally, this tagging system would be able to consolidate instrument specific morphologies within the same framework without producing conflicts. As an example, a source could be tagged as `compact' in a low resolution survey while having specific morphological features captured by tags referring to observations made by higher resolution instruments. 
This work aims to build the framework for such a tagging system for the first time.

The newest radio instruments in operation are producing maps of sources which are so deep, resolved, and with such high dynamic range that our existing classification schemes are failing. An updated, and extensible, radio morphology taxonomy of tags would be a tremendous benefit moving forward because deeper and wider surveys are expected to be a massive driver of scientific development in the coming decades. If the scientific community had a framework and terminology which were not intrinsically limited by sensitivity or resolution, it would mean that we could work with the same framework regardless of the technological improvements to the instruments in the field. 
We therefore expect this work could have major implications in various scientific contexts, including population studies and rare object searches in observations made by current and future radio instruments including 
the Australian Square Kilometre Array Pathfinder \citep[ASKAP;][]{Johnston2008ASKAP}, 
the Low Frequency Array \citep[LOFAR;][]{LOFAR2013},
the Deep Synopic Array 2000 \citep[DSA-2000;][]{Hallinan2021DSA2000}, 
the Murchison Widefield Array \citep[MWA;][]{Tingay2013MWA}, 
MeerKAT \citep[][]{Jonas2016MeerKAT}, 
the next generation Very Large Array \citep[ngVLA;][]{ngVLA2020AAS}
and the Square Kilometre Array \citep[SKA;][]{Dewdney2009SKA}. 

This work uses data from the Evolutionary Map of the Universe \citep[EMU;][]{Norris+2011}, a radio survey being conducted with the ASKAP telescope. ASKAP's large field of view means that it can map a large portion of the sky at once. Because of this, EMU is currently planned to map three quarters of the sky, the first two thirds of which are planned to be completed in the first five years. EMU is estimated to catalogue 40 million sources \citep[estimate made using the Tiered Radio Extragalactic Continuum Simulation method, T-RECS;][]{Bonaldi+2019}. In an effort to classify these sources at scale, we are launching `Radio Galaxy Zoo EMU' (RGZ EMU). RGZ EMU is a citizen science project designed to allow the public to provide valuable and essential insights into these sources, including host identification, source assembly, and source classification \rev{(full details on RGZ EMU in Tang, Vardoulaki, et al. in prep.)}. While designing this project, we discussed what classifications we would ask the citizen scientists to use. It became clear that there was no consensus on the terms to use. In part as a response to this dilemma, we collect plain English annotations on radio galaxies and implement a novel framework to derive semantic plain English tags.

The proposed process uniquely combines existing natural language processing (NLP) methods.
NLP has garnered significant research interest over the last twenty years \citep[see for instance][]{mishra2020natural}. \citet{Thomas2022MLforSciencePlanning} use NLP and a form of topic modelling called latent Dirichlet allocation
 \citep[LDA;][]{Vayansky2020TopicModellingReview} with the aim of guiding the planning process of research priorities by analysing trends in previous publications. \citet{Grezes2021astroBert} use deep learning based NLP techniques with the aim of improving the SAO/NASA Astrophysics Data System (ADS\footnote{See: \url{https://ui.adsabs.harvard.edu}}).

The method we present is not bound to radio astronomy. It can be applied to any domain. The code used in this work is publicly available at \url{https://github.com/mb010/Text2Tag} and is written to be transferable to other fields.

Our work is structured as follows. In Section~\ref{sec:data} we detail the data used in and collected through our experiments. We \rev{present the} proposed method \rev{in full in} Section~\ref{sec:method} before \rev{presenting the details of its application and the resulting taxonomy in Section~\ref{sec:taxonomy}. Initial physical results using the semantic taxonomy are presented in Section~\ref{sec:Initial Physical Results}.} The results and impact are discussed in Section~\ref{sec:discussion} and conclusions are made in Section~\ref{sec:conclusions}.
\nocite{scikit-learn}
\nocite{reback2020pandas}
\nocite{mckinney-proc-scipy-2010}
\nocite{Waskom2021}
\nocite{Hunter:2007}
\nocite{SciPyProceedings_11}
\nocite{spacy2}

\section{Data}
\label{sec:data}
Two experiments were designed and executed. Both made use of early versions of cutouts prepared for the RGZ EMU project as described in Section~\ref{subsec:Image Data}. The intent and design are detailed for the \textit{Plain English Annotations} experiment as well as the \textit{Expert Classification} experiment in Sections~\ref{subsec:Plain English Annotations} and \ref{subsec:Expert Classification} respectively. An anonymised version of the data is publicly available\footnote{\url{https://zenodo.org/record/7254123\#.Y7VvGtLP3Lp}}.

\subsection{Image Data}
\label{subsec:Image Data}
\begin{figure}
    \centering
    \includegraphics[width=\linewidth]{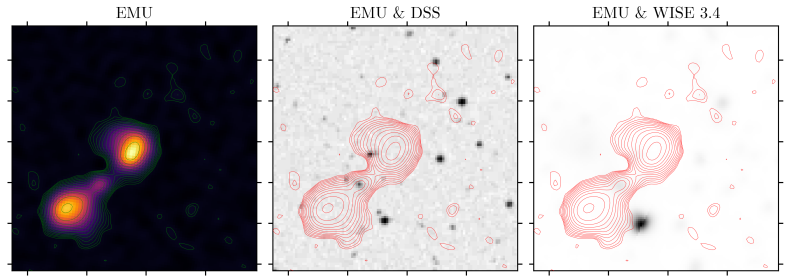}
    \caption{Example cutout as presented to the participants of the experiments detailed in Sections~\ref{subsec:Plain English Annotations} and \ref{subsec:Expert Classification}. Cutout centre: 21h\,02m\,16s -54$^\circ$\,$23^\prime$\,$36^{\prime\prime}$ (J2000). \rev{The lowest contour is fixed at 0.12\,mJy. Subsequent contours are multiples of $\sqrt{2}$ higher.}}
    \label{fig:cutout example}
\end{figure}

To produce the data analysed in this work, users were asked to consider individual images in turn. An example of one of these images is presented in Figure~\ref{fig:cutout example}. These images are early versions of the data to be used in the RGZ EMU project. This version consists of three panels containing a 6\,arcmin by 6\,arcmin cutout from the EMU pilot survey. The panels show EMU contours with the false colour EMU image, a Digitized Sky Survey \citep[DSS;][]{lasker1990guide} cutout, and a Wide-field Infrared Survey Explorer \citep[WISE;][]{Wright2010WISE} cutout. Each image is centred on an EMU Selavy catalogue component \citep[][]{Norris+2011,Norris+2021b}. 

The cutouts are subject to a number of criteria designed to select a small number of sources for early testing. Components which had an angular extent of less than 27\,arcsec (1.5 beam widths) were removed. Components which were within 45\,arcsec (2.5 beam widths) of another catalogued component were also removed, as these are largely simple doubles with little to no morphological features and can be classified algorithmically. 
\rev{This resulted in a list of 306 sources, for which cutout images were made. Our final sample consists of 299 of these cutouts because an undetected upload error caused seven cutouts to not be uploaded to the Zooniverse platform.}

\subsection{Plain English Annotations}
\label{subsec:Plain English Annotations}
To derive the desired plain English taxonomy, we started with plain English descriptions (annotations) of the given object or phenomenon. Using the 299 sets of images outlined in Section~\ref{subsec:Image Data}, we built a private Zooniverse\footnote{See: \url{https://www.zooniverse.org/}} project, where we presented users with an empty text box and a prompt reading:

\vspace{8pt}
\noindent\textit{Please describe the source:
\begin{itemize}
    \item in the \textbf{middle} of the frame and any \textbf{associated} emission.
    \item use \textbf{simple} English.  
    \item avoid \textbf{jargon}.
    \begin{itemize}
        \item e.\,g. refrain from typing FRI, WAT, etc.
    \end{itemize}
    \item descriptions should be \textbf{separated by ``,''}.
\end{itemize}}

This data was intentionally collected to be relatively unconstrained to encourage the annotations to cover diverse ideas of source features. Thus users were enabled to highlight and describe whatever caught their attention within the image. The trade-off in this unstructured description approach is that the resulting data are unwieldy and noisy (the consistency and formatting of phrases is not constrained). As such, the method outlined in Section~\ref{sec:method} contains a significant overhead of data cleaning which is common with any unstructured natural language data.

The data collection for this experiment ran from the 17$^{\rm th}$ of December 2021 to the 27$^{\rm th}$ of January 2022. We offered users who processed more than 100 sources co-authorship on this publication which is a direct result of their efforts. In total, we had 19 users annotate an average of 154 sources each, resulting in a total of 2,920 descriptions consisting of a \rev{8,486} comma separated annotations. Almost all of these users are astronomers, and more than three quarters of them have at least some academic experience of radio morphologies.

\subsection{Expert Classification}
\label{subsec:Expert Classification}
We conducted a second experiment to collect expert classifications on the same sets of images. This experiment was conducted with the aim of extracting ideas represented by annotations which are relevant to the expert's science cases.

We established a separate private Zooniverse project and invited a number of experts to participate in classifying the radio morphologies of the objects in the images described in Section~\ref{subsec:Image Data}. To classify objects with predefined classes participants were prompted with:

\vspace{8pt}
\noindent\textit{\textbf{Radio Morphology}: Please describe the source:
\begin{itemize}
    \item in the \textbf{middle} of the frame and any \textbf{associated} emission
    \item select one or more tags that fit object radio morphology.
\end{itemize}}

We presented the participants with 22 classes, which they could use as they wished, including assigning none or all of them to the subject. 
The abstract classes listed were selected from a compiled list of radio morphology classes presented in \citet{Rudnick+2021} and were: Single, Double, Classical double, Triple, Narrow-angle tail (NAT), Wide-angle tail (WAT), Bent tail, Fanaroff \& Riley Class 1 (FR I), Fanaroff \& Riley Class 2 (FR II), Fanaroff \& Riley Class 0 (FR 0), Hybrid, X-shaped, S-shaped, C-shaped, Diffuse, Double-double (DDRG), Core-dominant, Core-jet, Compact Symmetric Object (CSO), 1-sided, Odd Radio Circle (ORC), and Star-Forming Galaxy (SFG).
This experiment ran from the 27$^{\rm th}$ of January to the 19$^{\rm th}$ of May 2022. 5 experts made a total of 1,257 multi-label classifications of an average 251 objects each.

\section{Method}
\label{sec:method}
To the best of our knowledge, there is no existing process or NLP approach which produces a semantic taxonomy from a corpus of short annotations. The closest approaches are widely used topic modelling approaches. These approaches capture topics within a corpus through distributions of terms in documents. \citet{Vayansky2020TopicModellingReview} present a helpful review of topic modelling variants.
These models are designed to return a distribution of terms which belong to each discovered topic. They are not designed to return terms which communicate what a given topic is. 
We explicitly want to build a taxonomy on a certain subject. Therefore the terms which effectively capture the meaning of a topic are essential. 

Although a panel of experts may be able to manually define a semantic set of terms for a given problem, the success of such an approach would depend on whether the panel agree, the backgrounds of the experts, and their ability to distil complex ideas into simple plain English effectively. This manual approach would likely also lack the reproducibility and tractability that is expected by the physical sciences.

\begin{figure}
    \centering
    \includegraphics[width=\linewidth]{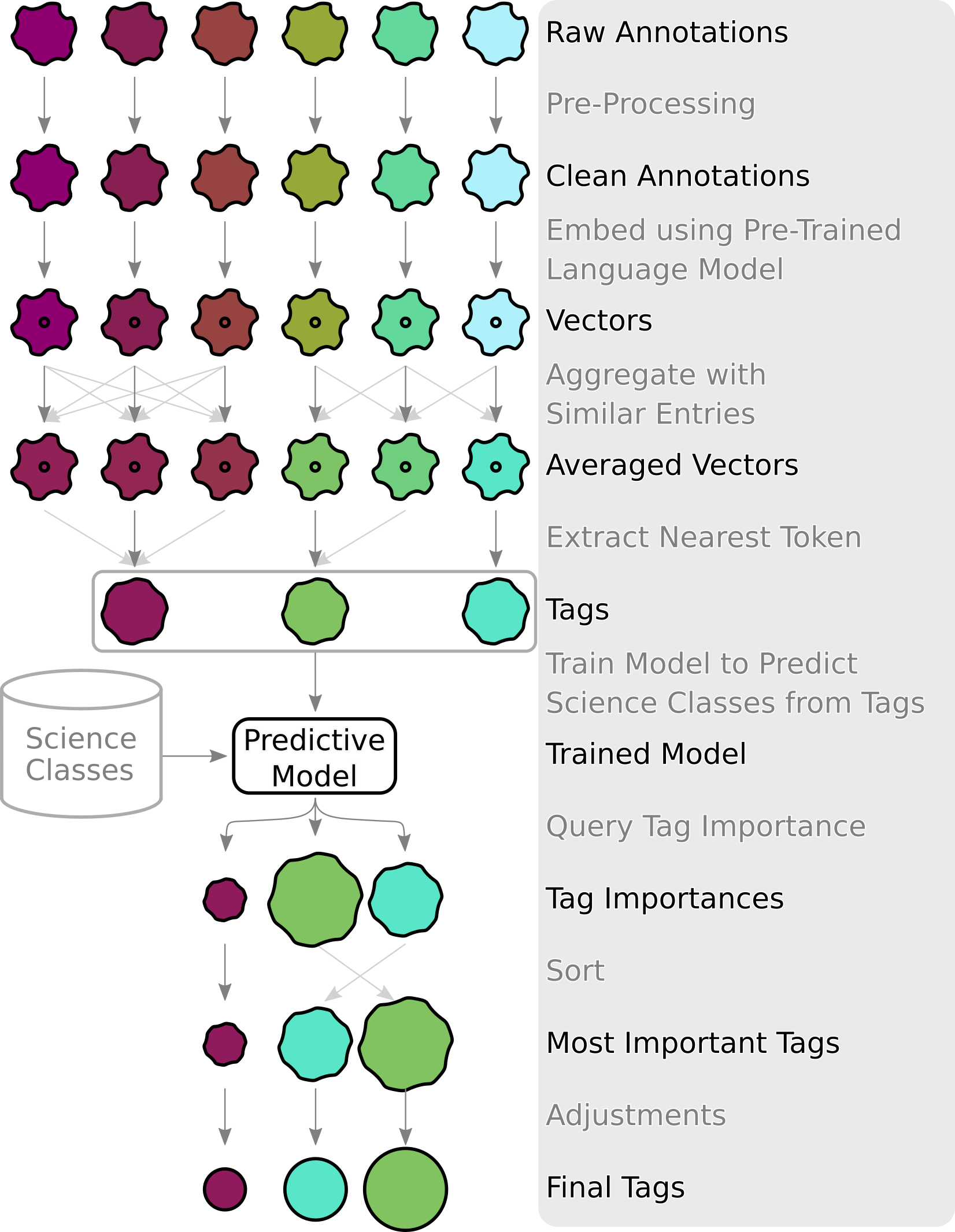}
    \caption{Proposed workflow to computationally derive a semantic plain English taxonomy from a set of annotations.}
    \label{fig:Text2Tags}
\end{figure}

We therefore propose a method through which short annotations are distilled into semantic tags in accordance with a specific science case and its respective features (including classes).
The workflow of the method is presented in Figure~\ref{fig:Text2Tags}.
The derived taxonomy should provide wide coverage of objects of interest, have the ability to distinguish features, be clear in what semantic feature it describes, and be appropriate to the science cases.
\nocite{cramerifabio}

\textit{Conceptually}, in the framework similar annotations are aggregated to produce a single term which we call `tag'. We rank how important tags are based on the impact they have in classifying the existing abstracted science classes. A selection is made on the most important tags to form a taxonomy.

A technical outline of the method is presented in Section~\ref{subsec:Method Overview}. The implementation details for our data and project are presented in Section~\ref{subsec:Implementation}.

\subsection{Technical Outline}\label{subsec:Method Overview}
Sequences of words are processed where $w_i$ represents the $i^{th}$ word of $N$ in a given annotation, $\mathbf{a}_j = \left(w_1, w_2, ..., w_N\right)$. Here $\mathbf{a}_j$ is the $j^{th}$ annotation of the $M$ annotations in our corpus. Note that in the NLP literature, the equivalent of annotations would be `documents'. This method is expected to work best on extremely short annotations (documents), where each annotation contains a single idea.
The annotations are embedded into a $k$-dimensional vector through a pre-trained model, $f_{\text{emb}}$:
\begin{equation}
   f_{\text{emb}}\left(\mathbf{a}_j\right) = \mathbf{v}_j \in \mathbb{R}^{k}. 
\end{equation}
This is currently implemented such that the order of the words does not affect the encoding (i.e. in a bag-of-words paradigm). We embed each word within an annotation through
\begin{equation}
    f_{\text{emb}}\left(\mathbf{a}_j\right) = \frac{1}{N}\sum_{i=1}^{N}f_{\text{emb}}\left(w_i\right).
\end{equation}

For pairs of annotations, $(i,j)\in[1,M]^2$, a similarity value is calculated using the cosine similarity, $g_{\text{cs sim}}:\mathbb{R}^k\xrightarrow[]{}[-1,1]$, which takes the dot product of two vectors scaled by the inverse of the product of the euclidean norms of those vectors:
\begin{equation}
    g_{\text{cs sim}}(\mathbf{v}_i, \mathbf{v}_j) = \frac{\mathbf{v}_i\cdot\mathbf{v}_j}{\lVert\mathbf{v}_i\rVert \lVert\mathbf{v}_j\rVert}.
\end{equation}
According to a similarity threshold, $\sigma$, $M$ averaged vectors are then calculated through:
\begin{equation}
    {v}_i^\prime =
\frac{1}{M^\prime} \sum_{j=1}^M
\begin{cases}
    \mathbf{v}_j  & \text{if\hspace{10pt}} g_{\text{sim}}(\mathbf{v}_i, \mathbf{v}_j) \geq \sigma \\
    0              & \text{else,}
\end{cases}
\end{equation}
where $M^\prime$ is the number of non-zero elements being summed over. The model, $f_{\rm emb}$, used to embed the annotations is then used to produce the token which is closest to $\mathbf{v}^\prime_i$:
\begin{equation}
    f_{\text{emb}}^{-1}(\mathbf{v}^\prime_j) \approx s_q,
\end{equation}
where $s_q$ is the $q^{th}$ entry of all $Q$ unique derived tags. As tags are derived from the $M$ annotations $Q\leq M$.

For each annotated object, we define $\mathbf{t} = (t_1, ..., t_Q)$ as a vector encoding of the tags, where $t_q$ is 1 if that tag was present in an annotation associated with that object, or 0 if that tag is not associated with it.
As each object has multiple individual annotations associated with it, it can be described through its derived tags $\mathbf{t} \in \{0,1\}^Q$.

We consider each science class $y$ in the set of science classes $Y$. Using the encoded tag vector, $\mathbf{t}$, for each object, we fit a model, $f_y:\{0,1\}^Q\xrightarrow[]{}\{0,1\}$, to predict the presence of each science class $y \in Y$. For each model, $f_y$, and tag representation, $t_q$, we calculate an importance
\begin{equation}
    f_{\text{Importance}}(f_y, t_q) = I_{(q,y)} \in \mathbb{R},
\end{equation}
where a larger value of $I_{(q,y)}$ means that $t_q$, and subsequently $s_q$, are more important to the classification output.

To recover the importance of the $q^{th}$ tag, we take an average across all models, $f_y$, for a given tag, $s_q$. We take the support weighted average of the importance of each model 
\begin{equation}
    I_q = \frac{1}{Q}\sum_{y\in Y}n_y I_{(q,y)}.
\end{equation}
Here, $n_y$ is the number of (positive) class $y$ entries. \rev{Note that other weightings may be preferable depending on the available data and purpose. For instance, if a multi-objective regression task were used instead to calculate $I_{(q,y)}$, then a uniform weighting across tasks may be more appropriate.} We normalise the importance values, $I_q$, such that
\begin{equation}
    \sum_{q=1}^{Q}I^\prime_{q} = 1, \text{\hspace{10pt}and\hspace{10pt}} I^\prime_{q} \in [0,1].
\end{equation}
Finally, the tags, $s_q$, are sorted by their $I^\prime_q$.
Although tags are all expected to have some non-zero $I^\prime_q$, a majority of the information is contained within the top tags. Additionally, the tags ranked lowest in this scheme are expected to be noisy (e.g. annotations which contain incorrect spellings, reference otherwise irrelevant features, or are only impactful on a given prediction through the random association of its small sample size). We set an importance threshold to select the top $Q^\prime<Q$ tags. These most important tags, consisting of $Q^\prime$ strings, $s\in S_{\text{Taxonomy}}$, is the derived taxonomy.

Some of the tags, $s$, may require clarification to allow the tag to be clear upon first reading. To do so the raw annotations which that tag was derived from are taken into consideration, in order to verify what it represents.

\subsection{Implementation}
\label{subsec:Implementation}
The exact implementation of the method outlined in Section~\ref{subsec:Method Overview} will depend on the data being used. The details for our implementation are as follows.

\subsubsection{Pre-Processing}\label{subsubsec:Pre-processing}
The goal of pre-processing the annotations is to format the data in a uniform manner without disrupting the content (i.\,e. standardising grammar, spelling, and formatting). To do this, a number of common NLP data processes are applied. These are applied in the order they are presented in.

All annotations are set to lower case and all accents are removed from characters. Ampersands and new line commands are removed or replaced as appropriate. Forward slash and full stop characters are replaced with commas as they are often observed to represent separate ideas, which our method assumes are comma separated. Double whitespaces are corrected and hyphens are removed.

Based on manual inspection, additional corrections are made to a number of annotations. These are 
spelling mistakes such as `copact' being corrected to `compact'. We then drop any annotations which mention `DSS', `WISE' or `optical' as these annotations are not expected to be a comment on the radio morphology, which is our target of interest. 

At this point, the sets of comma separated annotations are separated into individual annotations. Contractions are expanded.
A list of stopwords\footnote{Stopwords are a list of superfluous words which when applied to a text removes instances of those words from the text. Examples include `it' and `the'. A list of default stopwords can be found in the SpaCy source code under 
spacy\/lang\/en\/stop$\_$words.py.}
is extended to include `like', as well as scientific terms which the pipeline should not be affected by as they are both technical and not related to morphology. These additional stopwords include `emu', `galaxy', `galactic', `emission' and `source'. Terms for cardinal directions (`north', `south`, `east', `west') are also added to the stopwords since our focus is on the features themselves instead of their position relative to a specific source. This stopwords list is applied to the annotations.

We consider both lemmatized\footnote{Lemmatization is replacing a word with its root word, e.\,g. `apples' becomes `apple'.} and unlemmatized approaches to the data moving forward.
Each annotation has now been cleaned, and the data are largely consistently formatted.
%

\subsubsection{Embeddings}\label{subsubsec:Embeddings}
When embedding the cleaned annotations into vectors, we use SpaCy's large English language model\footnote{\url{https://spacy.io/models/en\#en_core_web_lg}} which is the largest available model within the SpaCy package (v3.3.0) that has tokens (largely words) embedded. It contains 685k embeddings. For a given vector, the model can return the closest embedded word(s). This feature is essential to our process, and is a key factor in the decision to use this model. Other advanced, such as transformer based models, can take word order into account, but do not have these token embeddings. In the implemented SpaCy version, the vector embeddings are learned through the GloVe algorithm introduced by \citet{pennington2014glove}. GloVe (Global Vectors for Word Representation) is an unsupervised representation learning algorithm which aims to embed words in a space which presents various desirable features. These features include semantic and linguistic similarity, which is advantageous for our use case.

\begin{figure}
    \centering
    \includegraphics[width=\linewidth]{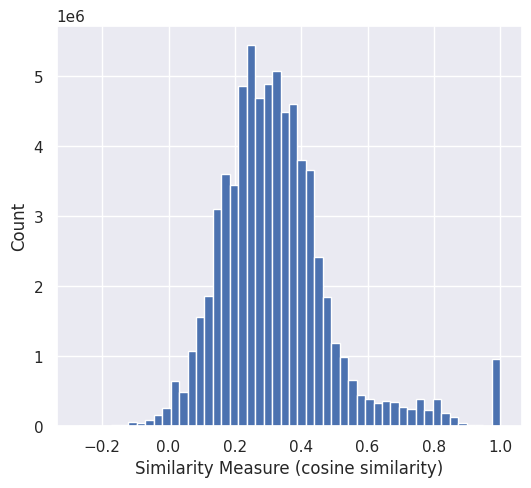}
    \caption{The histogram of the cosine similarities of the embedded annotations. Data is lemmatized and does not include self-similarity values.}
    \label{fig:similarities histogram}
\end{figure}

To decide what similarity threshold to use to aggregate over, we consider the histogram of cosine similarities of all annotations. This is presented in Figure~\ref{fig:similarities histogram}. This histogram is presented using annotation embeddings which were lemmatized and does not include self-similarities. The peak at 1.0 is due to tags which are identical (maximally similar). The rigorous cleaning process reducing short annotations to a few words before the annotations are embedded is likely a factor. Additionally, self-consistent vocabulary across multiple annotations may be embedded to an (essentially) identical vector. 

To capture the excess tail of more similar vectors, we consider similarity thresholds above $0.5$ for our models. Lower thresholds would contain the bulk of all annotation pairs, which would be counterproductive in trying to capture individual concepts. However, we would like to explore thresholds down to $0.5$ to reduce the number of unique terms and maximise the number of entries per unique tag. The tags derived from the aggregated vector encodings, are returned by SpaCy as single tokens (words).

\subsubsection{Model Definition}\label{subsubsec:Model definition}
We predict science classes on a source by source basis. We use 22 science classes to train the models to classify science classes from our derived tags. Three science classes presented to our expert classifiers are functionally removed as they are either not usable without spectra (\textit{Compact Symmetric Object\rev{; CSO}}), extremely contested and largely disused \citep[\textit{Fanaroff-Riley Class 0\rev{; FR0}},][]{Hardcastle2020, Rudnick+2021}, or largely uninformative when considering \rev{extended} radio morphology (\textit{Single}). Furthermore, the following three science classes have insufficient positive cases for training: Double-Double Radio Galaxies \citep[DDRG;][]{Schoenmakers+2000}, Hybrid Morphology Radio Sources \citep[HyMoRS;][]{Banfield+2015,kapinska+2017}, and Odd Radio Circles \citep[ORC;][]{Norris+2021a}.
The experts often did not agree with their usage of the terms. We explore what degree of agreement (expert threshold) beyond which we will consider a source as being positively classified with a certain science class in Section~\ref{subsec:Data config selection}.

To train models that predict science classes from our derived tags, we have 299 sources. This is a small data set. We chose a relatively simple model in response. We train random forests in a one-vs-rest scheme, i.\,e. 
one random forest model to classify one science class. We treat a set of these models as a single model which predicts the multi-label target of an input. The random forests use the Gini impurity criterion, with the aim of improving the explainability of the selected features \citep[][]{Menze2009}. The random forests are configured with 500 estimators, no maximum depth, and a seeded random state to allow for reproducible results. Unspecified features of the models are inherited from default values as implemented by Scikit-Learn v1.1.0.

\subsubsection{Evaluation}
Another challenge of the relatively small data set is the evaluation of the trained models. We use cross validation to maximise our use of the data. The model is evaluated through 10-fold cross validation, where we train 10 models on 10 different sets of nine tenths of the available data and evaluate each on the respectively withheld final tenth. With predictions for each tenth from one of the ten trained models, we recover predictions for each data point. These predictions result in approximate generalised performance metrics for the models.

We choose to track the performance of our models with macro and weighted F1 scores. An F1 score is the harmonic mean of precision and recall, and can be written as 
\begin{equation}
    F1 = 2 * \frac{(precision \cdot recall)}{(precision + recall)} = \frac{TP}{TP + \frac{1}{2}(FP + FN)},
\end{equation}
where $TP$ is the number of true positives, $FP$ is the number of false positives, and $FN$ is the number of false negatives.
The macro and weighted F1 scores are extensions to enable evaluations of multi-label problems. The macro F1 score is calculated by averaging F1 scores calculated in a one-vs-rest scheme for each target class. The weighted F1 score is calculated identically to the macro F1 score, except that we take the weighted mean where each score is first scaled by the number of samples of a given class, which may be more telling in an imbalanced data classification problem.

\subsubsection{Importance}
\label{subsubsec:Importance}
To estimate the importance of each of the tags, we use Shapley values. Shapley values are a common explainability tool used in machine learning applications \citep[][]{Lundberg2017SHAP}. These values convey how much a feature has contributed to the prediction of the respective model in comparison to the average predictions of the model. Exact Shapley values for each input tag and science classification are calculated using the trained random forest model and the \verb|SHAP| package for trees introduced in \citet{lundberg2020local2global}. 
These values are the importance values used to estimate which tags capture the semantics of radio morphology.

\rev{\section{Taxonomy}\label{sec:taxonomy}}

\subsection{Data Configuration}
\label{subsec:Data config selection}
To evaluate which data configuration (i.e. data processing parameter selection) is best, we consider the F1 scores of models trained on various configurations of the data. We grid search data across configurations including four expert thresholds, eleven similarity thresholds, and with or without lemmatization. This results in 88 configurations, which we construct and evaluate.

\begin{figure}
    \centering
    \includegraphics[width=\linewidth]{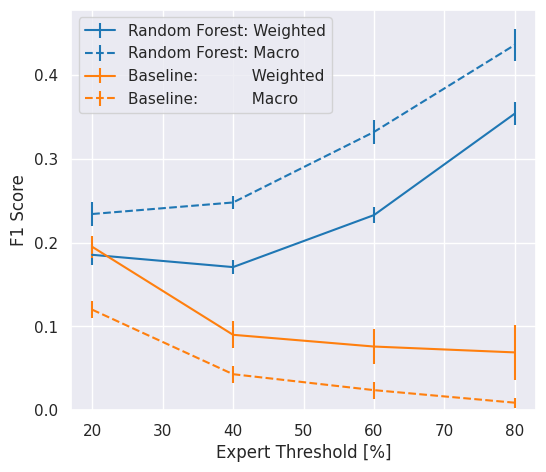}
    \caption{Aggregated F1 scores of both random forest models and baselines showing the effect of thresholding the expert classifications. The baselines are constructed by randomly predicting in accordance with the rate of positive cases per positive class. Statistics are measured over 5,000 instances of the baseline estimators, and over all lemmatization and similarity threshold configurations of the random forest models.}
    \label{fig:F1 scores expert thresholds}
\end{figure}

For the expert classification we demand that at least 20\,\%, 40\,\%, 60\,\%, or 80\,\% of the votes made on a given source agree. We call these confidence thresholds. Note that we make use of percentages. Sources which have not been classified by all experts can still have their classifications reflected in the confidence thresholds used in this search. We do not consider 100\% agreement amongst experts as so few classifications would survive that we could not train a model (highlighting a serious issue of the current classification scheme). The F1 scores are presented in Figure~\ref{fig:F1 scores expert thresholds}, for which the statistics of the random forest models are taken over all 22 configurations (2 lemmatization and 11 similarity threshold combinations) for each expert threshold. 

Simply stating that the model improves as the expert threshold is increased is largely true, see Figure~\ref{fig:F1 scores expert thresholds}. However, with increasing expert thresholds, the task which the model has been asked to complete becomes easier 
as the \textit{noise} in the classifications is functionally reduced. The subset of data where experts have a high consensus are more likely to have clearly identifiable morphologies (reduced aleatoric uncertainty) or present with a morphological classification which is more widely agreed upon amongst the experts (reduced epistemic uncertainty).
In an attempt to capture a broader perspective on what radio morphologies are, while maintaining accuracy of the classifications, we select an expert threshold of 60\% for the remainder of this work.

\begin{figure}
    \centering
    \includegraphics[width=\linewidth]{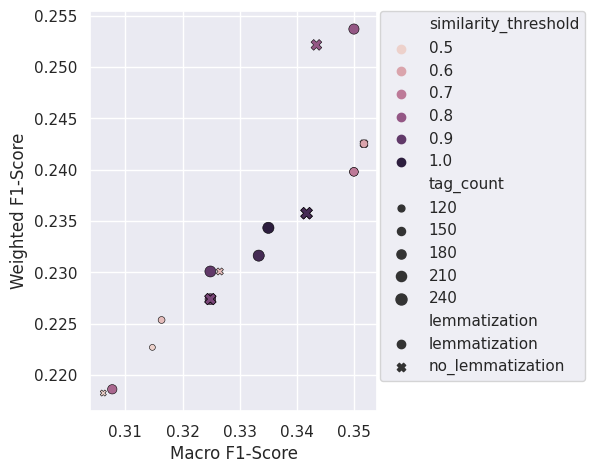}
    \caption{Grid search results across both lemmatization and the eleven similarity threshold configurations for the selected expert threshold of 60$\%$. tag\_count refers to the total number of unique tags which that configuration produced, which is strongly dependant on the similarity threshold of the configuration. The selected configuration is the lemmatized version with a similarity threshold of 0.80 (expert threshold of 60$\%$).}
    \label{fig:Grid Search}
\end{figure}

Figure~\ref{fig:Grid Search} shows the performance of the 22 models for each similarity threshold and lemmatization configuration. We select the configuration with a similarity threshold of 0.80 and lemmatized inputs. This configuration results in 213 unique tags. The model achieved a weighted F1 score of 0.254 and a macro F1 score of 0.350. This is the model with the highest weighted F1 score. The highest macro F1 score is 0.352 held by both configurations with a similarity threshold of 0.6.

\subsection{Tag Ranking}
\label{subsec:tag ranking}
The Shapley values are calculated for each tag provided to the model with respect to the model's outputs and the full data set. This provides us with a Shapley value for each science class and tag combination. We take the support weighted average of the Shapley values across the science cases to provide us with a descriptive Shapley value for a given tag. We normalise these values so that they sum to one across all tags. We call these values the comparative weighted Shapley values. These are presented as percentages which reflect how much sway a given tag has over the science classification of the model.

\begin{figure}
    \centering
    \includegraphics[width=0.98\linewidth]{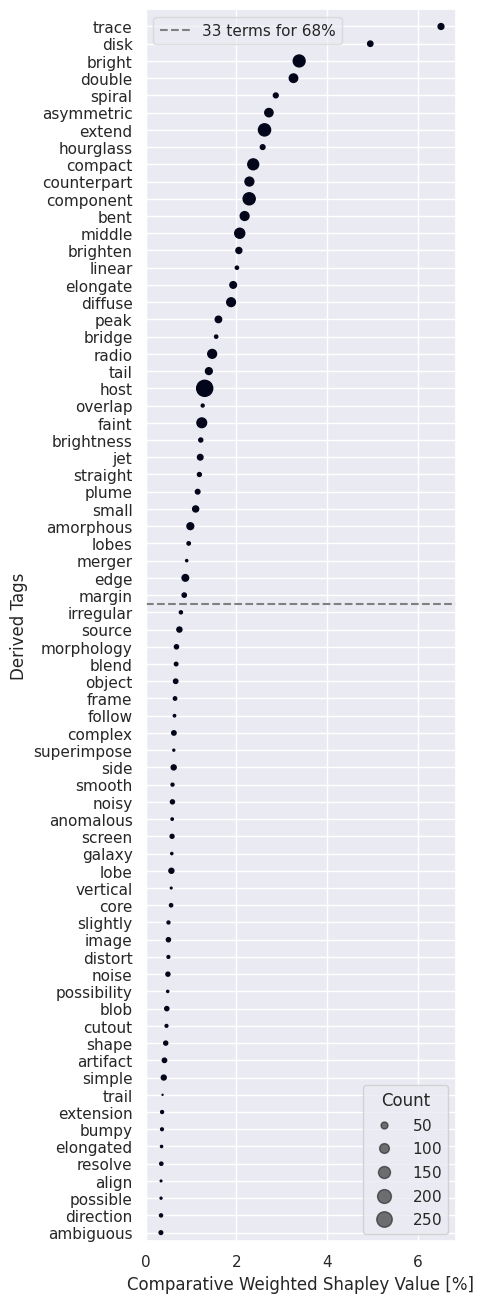}
    \caption{Top 70 tags sorted by their comparative weighted average Shapley values for the configuration selected in Section~\ref{subsec:Data config selection}. }
    \label{fig:Tag importance plot}
\end{figure}

We calculate the comparative weighted Shapley values and present the 70 most important terms in Figure~\ref{fig:Tag importance plot}. To select a usable volume of tags, we define the taxonomy to be the top tags required for 68\% of the descriptive power to be maintained (approximately 1\% of the comparative weighted Shapley value). In this data configuration, this results in a taxonomy of 33 tags.

\begin{figure}
    \centering
    \includegraphics[width=\linewidth]{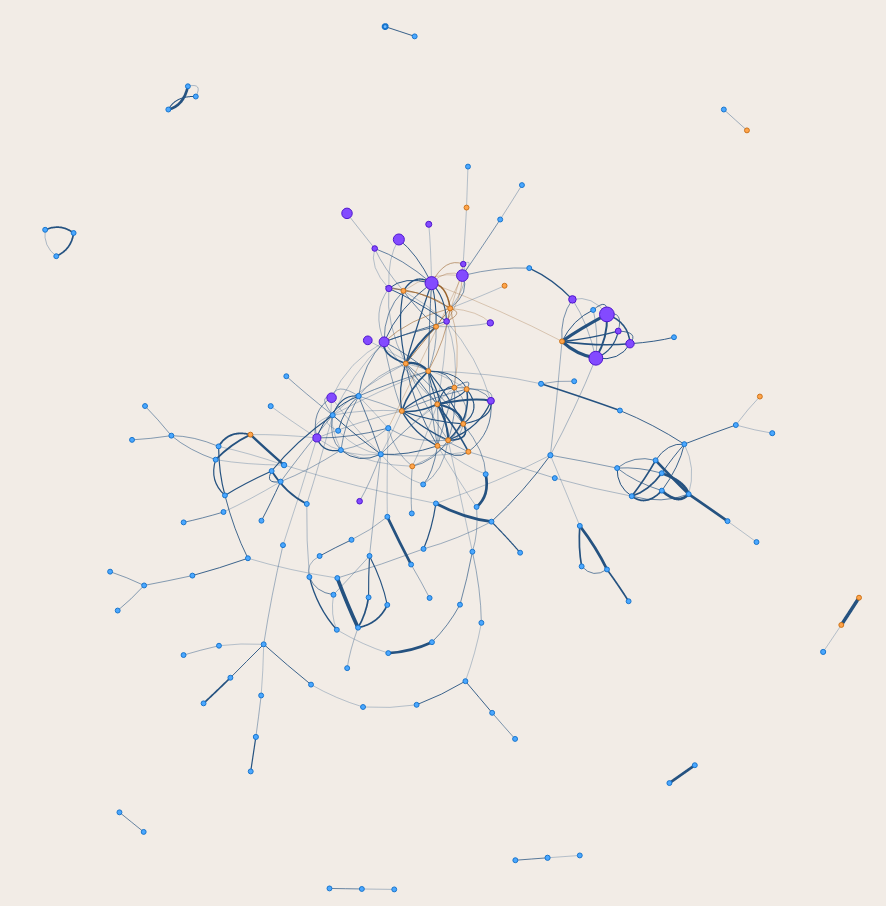}
    \caption{Still frame from the interactive graph available at \url{https://github.com/mb010/Text2Tag/blob/main/CorrelationsGraph.html} (download and open with a browser). The graph's edges are correlations above a magnitude of 0.3. Nodes are tags from the top top 33 (purple), other tags (blue), and science classes (orange). The node sizes reflect the occurrence rate of a given node across the 299 sources. The width of the edges represents the correlation between the two nodes it connects. All correlations are in the range $[-0.5,0.9]$.}
    \label{fig:Graph Screenshot}
\end{figure}

Highlighting the benefits of Shapley ranking over a simpler approach such as correlations, we present an interactive graph visualisation of moderately strong correlations between all combinations of both tags and science cases in Figure~\ref{fig:Graph Screenshot}. Importantly, the graph does not contain all 33 tags. This is because most of the top 33 tags are not strongly correlated with other terms. They are still the most impactful to the model's decision, as non-linear combinations of tags can be used to classify the science case. Their value to the non-linear classifications is captured by Shapley values.

\subsection{Taxonomy Adjustments}
\label{subsec:Taxonomy adjustments}
Limited to single words, the derived tags may not be an optimal selection. Suboptimal tag representations may also occur when the conjugation of a given term is relevant to the use case but lemmatization has removed it even if it would been more easily understood (e.g. `extended' in comparison to `extend').
Furthermore, the method only outputs single words, even if the concept the tag represents is better represented by multiple terms.

We therefore investigate each tag in turn by considering all annotations which contribute to it. We adjust tags in an attempt to optimise the taxonomy for grammatical and conceptual clarity. The adjusted tags are listed below, including descriptions of the original annotations / concepts which a tag represents. 
\begin{itemize}
    \item \textit{trace}: derives directly from numerous annotations stating `traces host galaxy'. 
    This is a more clear expression for what this tag represents. We therefore alter `trace' to `traces host galaxy'.
    
    \item \textit{disk}: derives from annotations such as `emission from galaxy disk'. We therefore merge this with `traces host galaxy' (originally `trace'). This refers to the radio emission tracing the host galaxy rather than the morphology of the host.
    
    \item \textit{bright}: refers to bright features of a presented cutout. This includes cores as well as neighbouring sources. This information is more clearly contained in the catalogues of radio component fluxes. Therefore, this tag is dropped.
    
    \item \textit{spiral}: derives from spiral galaxies being the hosts of the radio emission. This tag is changed to `traces host galaxy' as it then contains the relevant radio morphology information.
    
    \item \textit{asymmetric}: original annotations refer to asymmetric structure. To highlight the difference between asymmetric structure and brightness, we rename this tag to `asymmetric structure'.

    \item \textit{extend}: Grammatical adjustment to `extended'.
    
    \item \textit{component}: refers to the number of components which the source is composed of. This tag is dropped in favour of catalogues which list how many separated components are assigned to a source.
    
    \item \textit{counterpart}: refers to matching emission in either optical or infra-red. Host identification and `traces host galaxy' will capture this information. Therefore, this tag is dropped.
    
    \item \textit{middle}: Largely referring to presence and features of the central core of a radio galaxy. We therefore rename this tag to `core'.
    
    \item \textit{brighten}: Refers to `edge brightened' sources. We therefore clarify this by altering this tag to `edge brightened'.
    
    \item \textit{linear}: Refers to non bent radio morphologies, which is captured in the absence of the `bent' tag. Therefore this tag is dropped.
    
    \item \textit{elongate}: Refers to elongated structures in the radio emission. This is captured by the absence of the `bent' tag when sources are also `extended' and is therefore dropped.
    
    \item \textit{radio}: Annotations were written commenting on the radio emission in general ways (such as the presence of a jet or how many components are visible). This information is all mapped by other tags / processes. For this reason, we drop this tag from the taxonomy.
    
    \item \textit{overlap}: Often refers to emission which overlaps with radio contours or vice versa. It is therefore be changed to `traces host galaxy'.
    
    \item \textit{brightness}: Original annotations almost exclusively refer to `asymmetric brightness' across components or within the structure being discussed. We therefore clarify this tag by changing it to `asymmetric brightness'.
    
    \item \textit{straight}: Refers to the non-bent structure of the radio galaxy. We therefore drop it in favour of the absence of the tag `bent'.
    
    \item \textit{lobes}: We make a grammatical change to `lobe' with the intent to make this tag less ambiguous for future users.
    
    \item \textit{edge}: Highlighting clear edges of sources, as opposed to diffuse edges. This is largely equivalent to and is merged into `edge brightened'.
    
    \item \textit{margin}: The annotations from which this tag derives refer to the source extending beyond the margins of the cutout. This is being accounted for with updated cutouts, and is not morphologically relevant beyond the angular extent of a source, which is better presented in a catalogue format.
\end{itemize}

\subsection{Semantic Taxonomy}
\label{subsec:Semantic Taxonomy}
After the adjustments made to the tags in Section~\ref{subsec:Taxonomy adjustments}, we have 22 unique semantic tags. In alphabetical order, the semantic tags we propose to use for radio galaxy morphology are:
amorphous, asymmetric brightness, asymmetric structure, bent, bridge, compact, core, diffuse, double, edge brightened, extended, faint, host, hourglass, jet, lobe, merger, peak, plume, small, tail, and traces host galaxy.

\subsection{Effectively Assigning Tags}
\label{subsec:Tags for citizen science}
We have succeeded in deriving semantic tags for radio morphologies (see Section~\ref{subsec:Semantic Taxonomy}). However, for RGZ EMU and other citizen science approaches it is not effective to ask citizen scientists to use 22 tags. Terms would likely be ignored in a long list, and users would easily bottleneck on a small number of tags, neglecting the remainder of the taxonomy. This would be detrimental to both the science case and the user experience.

To improve the scientific results as well as the user experience, and to make the most effective use of the citizen scientists' time and energy, we consider which tags within the taxonomy can be most easily computed algorithmically at other stages of processing, e.g. `small' is easily calculated through the angular extent of the assembly mask for a given source. We aim for 10 tags which can be presented on a single screen to the citizen scientists. We consider each term in turn, and outline how each term might be assigned in Table~\ref{tab:Assigning tags}.
\begin{table}
\begin{tabular}{ll}
\hline
\multicolumn{2}{c}{Proposed for Algorithmic Assignment}                          \\ \hline
asymmetric brightness            & Integrated flux ratio between source sections.\\
asymmetric structure             & Symmetric components around host.             \\
compact                          & Angular extent of the components.             \\
diffuse                          & Proportion of assembly mask with emission.    \\
double                           & A ‘component’ number of two.                  \\
edge brightened                  & Relative radial brightness distribution.      \\
extended                         & Angular extent of the source.                 \\
faint                            & Integrated relative flux.                     \\
host                             & Whether or not a host is identifiable.        \\
peak                             & Peak within the assembly mask.                \\
small                            & Angular extent of assembly mask.              \\
traces host galaxy               & Assembly mask and host emission correlation.  \\ \hline
\multicolumn{2}{c}{Proposed for Tagging}                                         \\ \hline
\multicolumn{2}{l}{amorphous, bent, bridge, core, hourglass, jet, lobe, merger, plume, tail} \\ \hline
\end{tabular}
\caption{A summary of the suggestions on how the tags of the final adjusted taxonomy are to be assigned. `Assembly mask' refers to a mask derived from the source assembly process where multiple source components are grouped as (likely) belonging to a single source. `Tagging' refers to citizen science support, or calibrated and trained machine learning models.}
\label{tab:Assigning tags}
\end{table}

The tags which we believe are least easily computed will benefit the most from citizen scientist input. These are the tags which are presented as `proposed for tagging' in Table~\ref{tab:Assigning tags}. These ten tags are those which the RGZ EMU project will present to its citizen scientist volunteers.

\subsection{RGZ EMU Early Feedback}
\label{subsec:RGZ alpha test}
While working towards our final release of RGZ EMU, we asked a small group of 16 testers who have never worked on radio galaxy studies before (8 from China, 7 from Pakistan and 1 from Germany) for feedback on an early version of the tags terms provided by a beta version of the pipeline presented in this work. The tags presented to the testers were: bent, bridge, complex, diffuse, distorted, elongated, hourglass, jet, plume, tail. 

In general, our testers found most provided tags self explanatory. The main concern which the testers raised, was around the definition of three words they were not very familiar with: `plume', `tail' and `elongated'. We believe there are two main contributions to this phenomena:
\begin{enumerate}
    \item Our testers were all non-native English speakers, which is likely to explain their struggle with the meaning of `plume',
    \item The testers showed differences in their thinking around terms. For example, this included describing `elongated' as `extended', `tail of a comet', `oval shape', or a `jet-like structure'.
\end{enumerate}

To address these concerns, the final workflow \rev{will contain examples and conceptual definitions (see Appendix~\ref{appendix:definitions}) which users can reference for guidance.} Furthermore, the RGZ EMU team is considering the translation of the tags into multiple languages, where issues such as this may be less relevant.

\rev{
\section{Radio Astronomy Challenges and Semantic Morphologies
}\label{sec:Initial Physical Results}
}

\rev{
The semantic taxonomy derived in this work (Section~\ref{subsec:Semantic Taxonomy}) is expected to be most useful as a tool by which astronomers can  select samples of radio galaxies from source catalogues in a flexible manner. Assuming each of the tags is present or not, we can estimate how many populations can be selected. For the full taxonomy of 22 tags, $2^{22}=4\,194\,304$ populations can be selected ($2^{10}=1\,024$ for the ten tags that RGZ EMU citizen scientists will use; see Section~\ref{subsec:Tags for citizen science}). 
}

\rev{In practice, the number of populations that the tags map may be quite different. For example, the binary estimate presented here does not consider the use of other catalogued features, such as flux, spectral index, or redshift. It also does not take into account that certain tags may be fundamentally correlated. Additionally, one might expect that given enough data, catalogues containing vote fractions for each tag could enable uncertainty and strength estimates, i.e. how `bent' a source might be could be approximated by the fraction of citizen scientists which return the tag for that source.
}

\rev{
To demonstrate the utility of such semantically selected samples, we here synthesise a catalogue and perform some example selections. 
To synthesise our small data set into a catalogue, we estimate the tags for this catalogue by considering a source to have been assigned a given semantic tag if at least one of its annotations maps onto one of the tags in our final taxonomy. In this way, we treat our source annotations as a tagged catalogue\footnote{Available at: \url{https://github.com/mb010/Text2Tag/blob/main/data/mock_catalogue.csv}}. Future catalogues will likely improve upon this synthesised catalogue through multiple individuals making direct use of available semantic tags.
}

\rev{
We use this pseudo catalogue to demonstrate the impact that catalogues using a semantic taxonomy can have by considering two practical use cases. Firstly, we demonstrate the recovery of an existing population of radio galaxies in Section~\ref{subsec:Detecting traditinoal Populations}. We then highlight our ability to find morphological outliers in Section~\ref{subsec:Selecting abnormal sources}. 
}

\rev{
\subsection{Detecting Traditional Populations}\label{subsec:Detecting traditinoal Populations}
}

\rev{
We demonstrate how traditional populations can be recovered by recovering star-forming galaxies. We do this by considering sources tagged with \textit{traces host galaxy}. In practice we query our data for sources which were originally tagged with `trace'. This is the closest proxy to `traces host galaxy' tag that we can produce with our current data (see Sections~\ref{subsec:tag ranking} and \ref{subsec:Taxonomy adjustments}).}

\rev{By simply considering sources with the `trace' tag, we identify 38 objects. These are listed along side their respective expert star-forming galaxy (SFG) classification and estimated tags in Table~\ref{tab:sfg population}. This simple approach recovers 33 of the 45 sources labelled as SFGs by our experts (with at least 60$\%$ expert agreement).}
\begin{table*}
\begin{tabular}{cccl}
\hline
No. & \begin{tabular}[c]{@{}c@{}}Coordinates\\ (J2000)\end{tabular} & \begin{tabular}[c]{@{}c@{}}Expert SFG\\ Classification\end{tabular} & Tags                                                                                         \\ \hline
1   & 20h\,22m\,17s -55$^\circ42^\prime52^{\prime\prime}$  & \checkmark  & asymmetric structure, compact, double, faint, host, peak, \textbf{traces host galaxy}                 \\
2   & 20h\,22m\,31s -55$^\circ16^\prime45^{\prime\prime}$  & \checkmark  & amorphous, extended, host, \textbf{traces host galaxy}                                                \\
3   & 20h\,23m\,00s -54$^\circ59^\prime23^{\prime\prime}$  & \checkmark  & amorphous, bent, compact, core, host, \textbf{traces host galaxy}                                     \\
4   & 20h\,23m\,12s -53$^\circ55^\prime43^{\prime\prime}$  & \checkmark  & compact, diffuse, host, tail, \textbf{traces host galaxy}                                             \\
5   & 20h\,26m\,02s -55$^\circ36^\prime02^{\prime\prime}$  & \checkmark  & bent, bridge, double, faint, host, hourglass, peak, \textbf{traces host galaxy}                       \\
6   & 20h\,26m\,53s -53$^\circ56^\prime33^{\prime\prime}$  & \checkmark  & amorphous, diffuse, host, \textbf{traces host galaxy}                                                 \\
7   & 20h\,29m\,31s -56$^\circ44^\prime34^{\prime\prime}$  & \checkmark  & amorphous, extended, host, \textbf{traces host galaxy}                                                \\
8   & 20h\,29m\,44s -57$^\circ57^\prime21^{\prime\prime}$  & \checkmark  & bridge, core, double, extended, faint, host, peak, \textbf{traces host galaxy}                        \\
9   & 20h\,31m\,29s -53$^\circ44^\prime17^{\prime\prime}$  & \checkmark  & amorphous, extended, faint, host, peak, \textbf{traces host galaxy}                                   \\
10  & 20h\,31m\,52s -53$^\circ46^\prime30^{\prime\prime}$  & \checkmark  & amorphous, compact, core, extended, host, merger, peak, \textbf{traces host galaxy}                   \\
11  & 20h\,32m\,02s -53$^\circ37^\prime56^{\prime\prime}$  & \checkmark  & amorphous, compact, extended, host, merger, peak, \textbf{traces host galaxy}                         \\
12  & 20h\,34m\,02s -52$^\circ58^\prime50^{\prime\prime}$  & \checkmark  & diffuse, extended, faint, host, \textbf{traces host galaxy}                                           \\
13  & 20h\,35m\,33s -57$^\circ22^\prime44^{\prime\prime}$  & \checkmark  & amorphous, compact, extended, faint, host, \textbf{traces host galaxy}                                \\
14  & 20h\,36m\,11s -57$^\circ09^\prime38^{\prime\prime}$  & \checkmark  & compact, core, extended, faint, host, peak, \textbf{traces host galaxy}                               \\
15  & 20h\,38m\,19s -54$^\circ05^\prime49^{\prime\prime}$  & \checkmark  & host, merger, peak, small, \textbf{traces host galaxy}                                                \\
16  & 20h\,40m\,18s -55$^\circ16^\prime18^{\prime\prime}$  & \checkmark  & compact, diffuse, extended, host, \textbf{traces host galaxy}                                         \\
17  & 20h\,40m\,36s -53$^\circ15^\prime53^{\prime\prime}$  & \checkmark  & bent, bridge, extended, host, merger, \textbf{traces host galaxy}                                     \\
18  & 20h\,41m\,09s -55$^\circ28^\prime19^{\prime\prime}$  & \checkmark  & compact, double, extended, host, \textbf{traces host galaxy}                                          \\
19  & 20h\,43m\,10s -53$^\circ27^\prime55^{\prime\prime}$  & \checkmark  & compact, faint, host, peak, small, tail, \textbf{traces host galaxy}                                  \\
20  & 20h\,43m\,41s -57$^\circ02^\prime09^{\prime\prime}$  & \checkmark  & diffuse, extended, host, \textbf{traces host galaxy}                                                  \\
21  & 20h\,43m\,55s -57$^\circ20^\prime04^{\prime\prime}$  & \checkmark  & amorphous, compact, core, extended, faint, host, merger, peak, \textbf{traces host galaxy}            \\
22  & 20h\,50m\,42s -55$^\circ47^\prime58^{\prime\prime}$  & \checkmark  & diffuse, extended, faint, host, \textbf{traces host galaxy}                                           \\
23  & 20h\,53m\,05s -56$^\circ25^\prime08^{\prime\prime}$  & \checkmark  & amorphous, compact, extended, host, \textbf{traces host galaxy}                                       \\
24  & 20h\,53m\,43s -54$^\circ02^\prime26^{\prime\prime}$  & \checkmark  & compact, extended, faint, host, peak, \textbf{traces host galaxy}                                     \\
25  & 20h\,55m\,25s -54$^\circ45^\prime40^{\prime\prime}$  & \checkmark  & amorphous, asymmetric structure, compact, extended, host, \textbf{traces host galaxy}                 \\
26  & 20h\,55m\,35s -55$^\circ05^\prime54^{\prime\prime}$  & \checkmark  & bent, core, diffuse, extended, host, lobe, peak, tail, \textbf{traces host galaxy}                    \\
27  & 20h\,59m\,43s -53$^\circ58^\prime52^{\prime\prime}$  & \checkmark  & amorphous, compact, extended, host, \textbf{traces host galaxy}                                       \\
28  & 20h\,59m\,56s -55$^\circ33^\prime47^{\prime\prime}$  & \checkmark  & diffuse, double, edge brightened, extended, faint, host, hourglass, peak, \textbf{traces host galaxy} \\
29  & 21h\,00m\,39s -54$^\circ29^\prime13^{\prime\prime}$  & \checkmark  & amorphous, compact, core, host, peak, \textbf{traces host galaxy}                                     \\
30  & 21h\,01m\,13s -57$^\circ14^\prime26^{\prime\prime}$  & \checkmark  & compact, extended, faint, host, peak, small, \textbf{traces host galaxy}                              \\
31  & 21h\,01m\,49s -57$^\circ56^\prime45^{\prime\prime}$  & \checkmark  & amorphous, compact, diffuse, host, \textbf{traces host galaxy}                                        \\
32  & 21h\,02m\,10s -55$^\circ04^\prime42^{\prime\prime}$  & \checkmark  & core, diffuse, extended, host, peak, \textbf{traces host galaxy}                                      \\
33  & 21h\,06m\,19s -56$^\circ48^\prime27^{\prime\prime}$  & \checkmark  & asymmetric brightness, compact, diffuse, extended, host, \textbf{traces host galaxy}                  \\
34  & 20h\,22m\,36s -56$^\circ16^\prime25^{\prime\prime}$  & $\times$    & asymmetric structure, compact, double, faint, host, peak, small, tail, \textbf{traces host galaxy}    \\
35  & 20h\,24m\,45s -56$^\circ20^\prime47^{\prime\prime}$  & $\times$    & asymmetric structure, compact, faint, host, small, \textbf{traces host galaxy}                        \\
36  & 20h\,47m\,46s -56$^\circ44^\prime04^{\prime\prime}$  & $\times$    & asymmetric structure, compact, core, diffuse, extended, host, tail, \textbf{traces host galaxy}       \\
37  & 20h\,58m\,37s -57$^\circ56^\prime39^{\prime\prime}$  & $\times$    & compact, host, \textbf{traces host galaxy}                                                            \\
38  & 20h\,59m\,51s -57$^\circ51^\prime21^{\prime\prime}$  & $\times$    & bent, compact, core, double, extended, faint, host, peak, small, \textbf{traces host galaxy}          \\ \hline
\end{tabular}
\caption{\rev{Sources selected for the \textit{traces host galaxy} tag (`trace' in practice; see Section~\ref{subsec:Detecting traditinoal Populations}). Using an expert threshold of $60\%$ (see Section~\ref{subsec:Data config selection}) we confirm whether or not the sampled sources are SFGs. For each source, the tags are listed in alphabetical order.}}
\label{tab:sfg population}
\end{table*}

\begin{figure*}
    \centering
    \includegraphics[width=\linewidth]{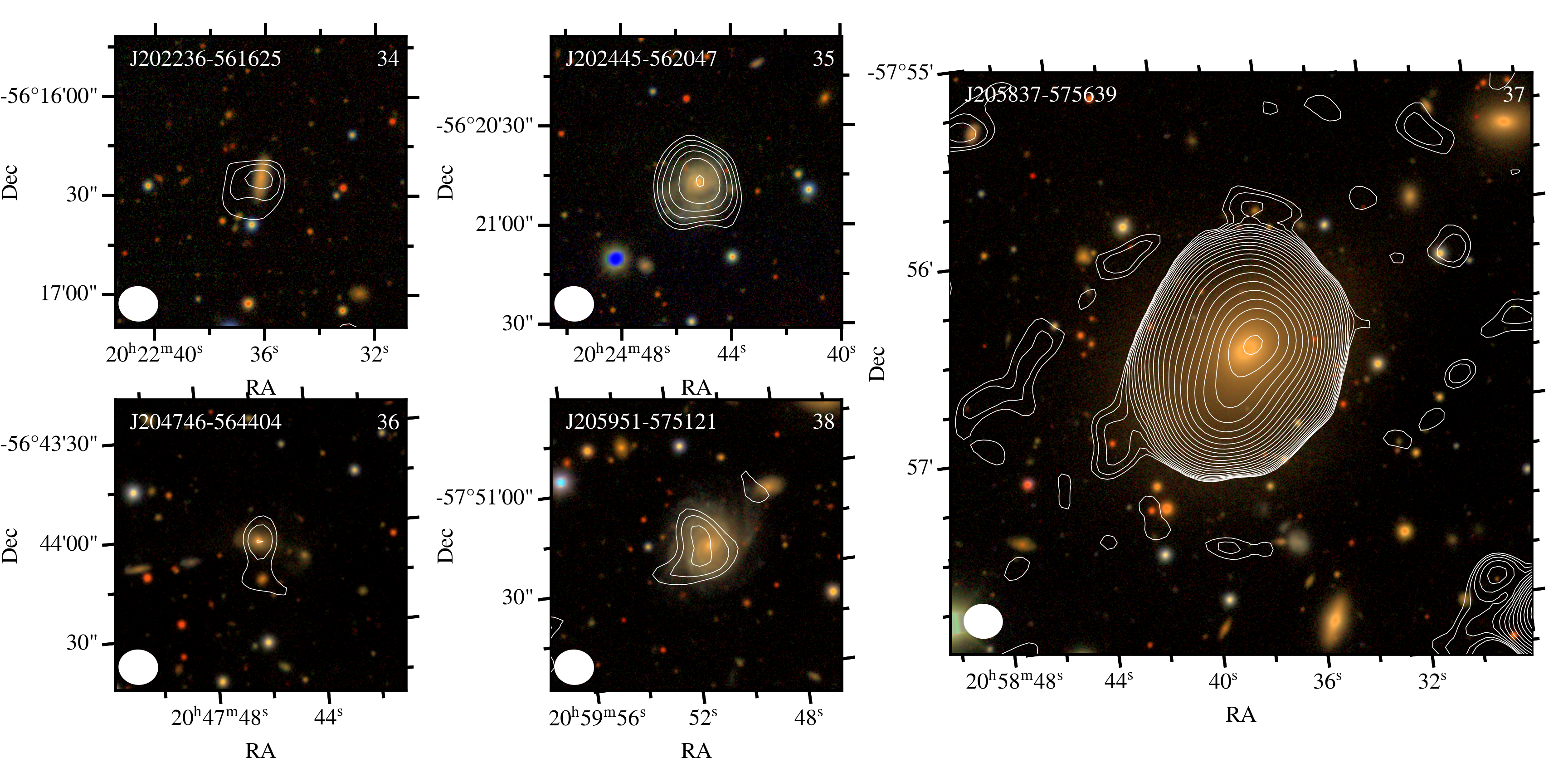}
    \caption{Sources 34-38 from Table~\ref{tab:sfg population} tagged with \textit{trace} (selected as a proxy for `traces host galaxy') which are not classified as SFGs by experts. EMU radio brightness contours matching those in Figure~\ref{fig:cutout example} with DES cutouts, combining g, r, and i band data into an RGB image following \citet[][]{lupton2004preparing}. Sources are annotated with their source numbers and (J2000) coordinates as presented in Table \ref{tab:sfg population}. Each source is shown to the same angular scale, as highlighted by the radio beam size in the bottom left of each panel.}
    \label{fig:SFG_composite}
\end{figure*}

\rev{
Five sources with the `trace' tag were not classified as SFGs in our expert classification scheme, i.e. 34 -- 38 
in Table~\ref{tab:sfg population}. Images of these sources are presented in Figure~\ref{fig:SFG_composite}, where their radio contours are shown overlaid on optical data from the Dark Energy Survey \citep[DES;][]{DES2018}. Combining the deeper and higher resolution DES optical data into an RGB image aids visual interpretation compared to using DSS greyscale images. Now the primary choice for the EMU Zoo project, DES data were not initially used due to concerns about accessibility and coverage. Each of these sources is discussed individually below with respect to the optical morphology catalogues presented in Walmsley et al. (in prep.) made with the Zoobot\footnote{\url{https://github.com/mwalmsley/zoobot}} package \citep[as described in][]{WalmsleyDECALSgz2022}, where the percentage of people who would have answered with a given feature is stated behind each catalogued feature.}

\rev{Source~34 is a smooth ($78\%$) cigar shaped ($70\%$) galaxy (could be an edge-on galaxy). 
Source~35 is a featured ($82\%$) face-on ($98\%$; not edge-on) spiral ($73\%$) galaxy without a bar ($70\%$). 
Due to selection cuts, Source~36 was not included in the Walmsley et al. (in prep.) catalogues; however,  the radio emission is expected to stem from at least two small galaxies bounded by the contours in the image.
Source~37 is a smooth ($67\%$) round ($59\%$) galaxy.
Source~38 is a featured ($91\%$) face-on ($98\%$; not edge-on) spiral ($98\%$) galaxy. It does not have a bar ($71\%$) but has a small bulge ($80\%$) and tightly wound spiral arms ($83\%$).
}
\rev{Consequently, of the five sources initially not classified as SFGs, when reconsidered with the deeper DES \citep[][]{DES2018} images and optical morphology catalogues, it is clear that at least two sources (35 and 38) are star-forming spiral galaxies. It is likely that the experts simply did not feel confident in classifying these sources as SFG with the limited resolution and sensitivity of the DSS data.
}

\rev{
Twelve additional sources that were classified as SFGs with 60\% agreement amongst our experts, but that \textit{did not} have the `trace' tag assigned to them are presented in Table~\ref{tab:sfg no trace tag}. If the selection had been made on `traces host galaxy`, rather than `trace', then these sources would have been included in Table~\ref{tab:sfg population}. This is because `traces host galaxy' is derived through multiple tokens such as `counterpart' (see Section~\ref{subsec:Taxonomy adjustments}). 
}
\begin{table*}
\begin{tabular}{ccl}
\hline
No. & \begin{tabular}[c]{@{}c@{}}Coordinates\\ (J2000)\end{tabular} & Tags \\ \hline
1   & 20h\,32m\,03s -53$^\circ44^\prime35^{\prime\prime}$ & amorphous, compact, core, extended, host, \textbf{traces host galaxy} \\
2   & 20h\,34m\,13s -54$^\circ01^\prime30^{\prime\prime}$ & asymmetric structure, compact, core, faint, host, tail, \textbf{traces host galaxy} \\
3   & 20h\,36m\,30s -57$^\circ07^\prime19^{\prime\prime}$ & asymmetric structure, compact, extended, host, small, \textbf{traces host galaxy} \\
4   & 20h\,44m\,36s -57$^\circ37^\prime29^{\prime\prime}$ & asymmetric structure, compact, extended, faint, host, small, tail, \textbf{traces host galaxy} \\
5   & 20h\,32m\,59s -55$^\circ38^\prime04^{\prime\prime}$ & compact, host, lobe \\
6   & 20h\,33m\,34s -54$^\circ31^\prime22^{\prime\prime}$ & amorphous, compact, extended, host, merger, peak \\
7   & 20h\,34m\,49s -54$^\circ12^\prime39^{\prime\prime}$ & amorphous, compact, diffuse, double, extended, host, hourglass \\
8   & 20h\,41m\,41s -56$^\circ10^\prime21^{\prime\prime}$ & amorphous, compact, double, faint, host, merger, peak \\
9   & 20h\,46m\,26s -54$^\circ00^\prime51^{\prime\prime}$ & diffuse, double, host \\
10  & 20h\,55m\,12s -54$^\circ31^\prime25^{\prime\prime}$ & amorphous, compact, extended, host, merger, small \\
11  & 20h\,56m\,44s -56$^\circ37^\prime48^{\prime\prime}$ & compact, diffuse, double, extended, host, merger \\
12  & 21h\,02m\,57s -54$^\circ29^\prime35^{\prime\prime}$ & amorphous, asymmetric structure, bent, core, diffuse, faint, host, peak, tail \\ \hline
\end{tabular}
\caption{\rev{Sources which had expert SFG classifications (above 60\% agreement) but were not selected through `trace' as described in Section~\ref{subsec:Detecting traditinoal Populations} and used for Table~\ref{tab:sfg population}.}}\label{tab:sfg no trace tag}
\end{table*}

With a consistent use of the `traces host galaxy' tag, the remaining eight SFG sources from Table~\ref{tab:sfg no trace tag} (sources 5 -- 12) are likely to be tagged as such, as their radio emission do largely trace the respective optical host (see Appendix~\ref{appendix:SFG no tag}).

\rev{
Of the 45 sources our experts classified as SFGs, the selection of \textit{traces host galaxy} would have recovered at least 82\% (33 and 4 from Tables~\ref{tab:sfg population} and \ref{tab:sfg no trace tag} respectively) of these sources, making it a strong candidate to select SFG populations from future catalogues and highlighting how such a catalogue can be used.
}

\rev{
\subsection{Rare Source Detection}\label{subsec:Selecting abnormal sources}
}

\begin{figure}
    \centering
    \includegraphics[width=\columnwidth]{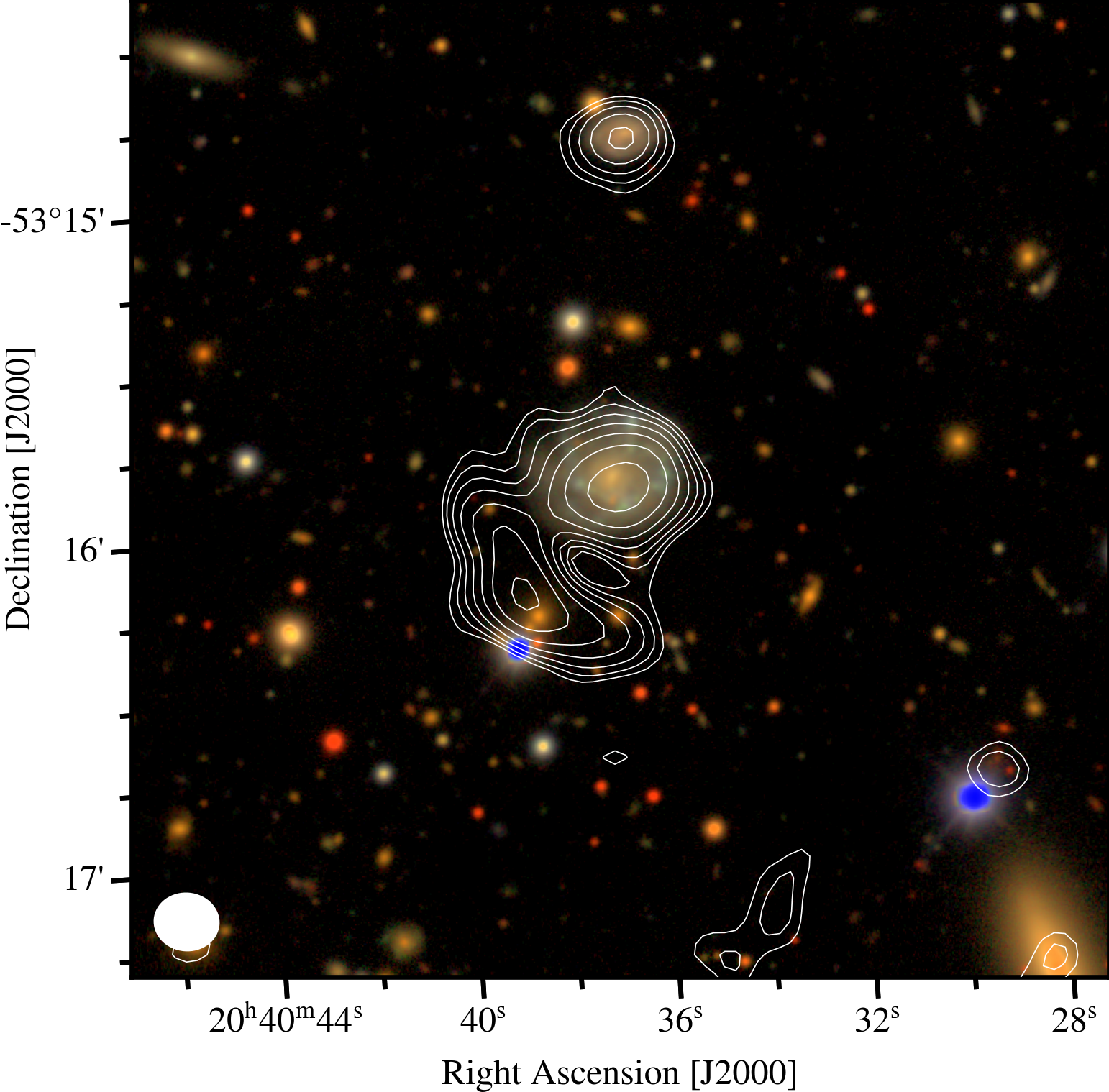}
    \caption{\rev{Rare source selected from the synthesised initial semantic tag catalogue by querying \textit{``hourglass $\setminus$ (amorphous $\cup$ traces host galaxy $\cup$ bent)''} (set theory notation). EMU radio brightness contours matching those in Figure~\ref{fig:cutout example} ontop of a DES cutout prepared as in Figure~\ref{fig:SFG_composite}.}}
    \label{fig:sourceB}
\end{figure}

We here highlight the flexibility and practicality of our taxonomy by considering a combination of tags that a radio astronomer might consider to be abnormal. We query to find a source which \rev{ a)} appears to be a merger, \rev{ b)} presents bridged features and \rev{ c)} is not faint. The result is a single entry: \rev{source 17 from Table~\ref{tab:sfg population}, shown in Figure~\ref{fig:sourceB}.}

\rev{This source is, as expected, an unusual object requiring expert followup.  It is a composite of emission from a flocculant spiral face-on 2MASS galaxy \citep{2006AJ....131.1163S}, plus apparently associated emission to its SE with no obvious separate optical counterpart.} 
\rev{
The burned out (blue) object at the southern edge of the contours is listed as two stars in the Gaia catalogues \citep{Gaia2016, Gaia2021}, and has no obvious connection to the radio structure. A very careful evaluation of the chances for serendipity, and the possible physical nature of this source, are beyond the scope of this paper.  However, the control provided by these semantic tags has allowed for the selection of an unusual object worthy of further study.
\nocite{Aladin2000,AladinLite2014}
}

\section{Discussion and Impact}\label{sec:discussion}

\rev{\subsection{Taxonomy}\label{subsec:Discussion on Proposed taxonomy}}

\rev{The proposed semantic tags will find immediate use in the RGZ EMU citizen science project (Tang $\&$ Vardoulaki et al. in prep.).} For future implementations of science tags being assigned to sources, we suggest that the community uses a hash symbol to denote the use of a tag (e.g. `\#compact') as suggested in \citet{Rudnick+2021} to distinguish from traditional classification frameworks. This should prove useful to the legibility and analysis of future catalogues and works.

This is the \textit{first step of the tagging framework} in radio astronomy morphology. The taxonomy is intentionally designed to be extensible, such that when the community decides a feature of interest is not being captured by the current version of the taxonomy \rev{it can be updated to include the appropriate tag. The presented set of semantic tags are a first step towards mapping common features of radio morphologies using plain English annotations.}

The \textit{specificity} of the tags in comparison to the current classification scheme may be a concern to some astronomers. The inconsistency with which current radio morphological classifications are defined means that the language currently in use does not have the desired specificity either - regardless of how specific a term is in an individual astronomer's mind. Furthermore, we highlight that terms are expected to be used more consistently and clearly when selected directly rather than being derived through annotations. 

A \textit{science class mapping} using the semantic taxonomy is one of the goals of the RGZ EMU team. This mapping will be constructed \rev{towards the end of the RGZ EMU project. This mapping should be able to provide the traditional classification of objects by predicting them based on the tags that citizen scientists have assigned to objects.} It will include the most common radio morphology science classes which the community is more accustomed to. While it is hoped that the tags will span the full space of possible scientific classifications, there may be cases where the provided mapping does not cover a science case perfectly \rev{in its current form.}

\rev{
Regardless, the ability to combine the tags to select semantic populations, as demonstrated in Section~\ref{sec:Initial Physical Results}, will enable feature specific population studies and  sources to be omitted if they present features that are not relevant to a given science case.
}
\rev{
\subsection{Experiment and Implementation}\label{subsec:Discussion on experiment and implementation}
}
\rev{As described, the experimental set up of this work has a number of limitations, including the small size of the data set: given the degree of  variation present across radio galaxy morphologies it is not possible to capture all abstract science classes of radio galaxies completely. For example, as stated in Section~\ref{subsubsec:Model definition}, ORCs are so rare that we could not train on them, and thus they are not taken into consideration in the weighted Shapley values. Such biases are therefore currently passed onto the derived and proposed semantic taxonomy.}

\rev{Additionally,} abstract science classes based primarily on morphology do not directly encode other physically relevant information. Given sufficient information, it may be more informative to derive semantic tags directly from physical parameters, e.g. active galactic nuclei accretion rates. \rev{This would encourage the derived semantic tags to carry information regarding the physics itself, rather than a proxy, e.g. abstract science classes.} In practice, collecting a large enough sample with which to do this is not feasible at this time, but should be considered in future approaches \rev{and other domains}.

\rev{We use a \textit{pre-trained NLP model} in our approach to this task. This model is a limitation even though it is also a key factor in the success of our implementation and experiments. For instance, the model we used only returns individual `tokens', and not fully grammatically correct terms or phrases. We amend this through manual inspection and adjustments, however we recognise this is not a scalable solution, and will not be possible in all situations. We hope that future approaches will find solutions to this problem. The NLP literature is currently developing at a significant pace. We therefore urge future iterations and applications of this approach to actively re-consider which pre-trained model is used. We expect that by using the most up to date pre-trained model, future implementations will have more robust and versatile encodings of annotations and tags.}

Finally, we note that the semantic radio morphology taxonomy derived in this work is inherently bound to the instrument with which the radio galaxies were imaged (ASKAP). Data from a more sensitive \citep[e.g. SKA;][]{Dewdney2009SKA} or higher resolution \citep[e.g. LOFAR long baseline][]{Morabito2022LOFAR_VLBI} instrument might require additional or different semantic tags. This would be simple to implement under the proposed tagging paradigm, as it would \rev{be sufficient to supplement} the existing taxonomy with the appropriate semantic tags.

\rev{\subsection{Semantic Taxonomies}\label{subsec:discussion semantic taxonomies}}

\rev{The proposed method, and respective task of deriving semantically meaningful tags \citep[first outlined in][]{Bowles2022_SemanticTags}, have the potential to impact other fields. We therefore discuss their potential impact and limitations in a domain agnostic tone.}
The move away from technical classes to \rev{semantic language capturing features} may have a broad impact across \rev{a number of technical fields}, especially where complex classes have been defined and the field has \rev{since} moved beyond those initially valuable classification schemes, as is the case in radio astronomy. This could include any feature rich data product, especially where features are often repeated across classes.

The \textit{collaborative} nature of science may be improved by the use of simplified \rev{and semantic} language. Complex ideas are often shrouded in equally complex terminology, which can be highly effective when experts communicate with one another, but quickly becomes a hindrance to communicating in any other situation. Capturing features of an object using dictionary level definitions will lower the barrier to entry for established researchers  \rev{who are not domain experts} to study \rev{the features captured by the semantic language}. This can \rev{be a significant benefit,} where domain specific terminology \rev{could} be \rev{an active} barrier to communication \rev{in inter-disciplinary research}. Additionally, \rev{the use of} plain English should enable scientific collaborations \rev{within a given field, i.e. between radio astronomy domain experts and astronomers who are not experts in radio morphology.}

\textit{Outreach} efforts \rev{are} also \rev{likely to} benefit from the change to language. We hope that the simple language will reduce the barrier to entry for those who would like to become \rev{experts in the respective field}. This will have direct impact on the accessibility of technical fields as a whole including communities who have not had much practice in the use of scientific language. This is in perfect alignment with the educational aspects of citizen science, which are often used to engage underprivileged communities in science with the aim to inspire and empower. The hope of citizen science outreach is that students who have seen, interacted with, and \rev{subsequently} added to the international body of science feel empowered to pursue STEM subjects. Clearer language will improve engagement to support this goal.

The science in \textit{citizen science} projects is also expected to benefit from the new language. Easily understood concepts presented by the simplified language should lead to improved usage of tags for a given source \citep[][]{Wald2016}.
\rev{Additionally, the reduction in training time / effort of the citizen scientists is hoped to lower the labelling cost for projects as a whole by reducing the labelling cost for individual citizen scientists.}

\textit{Deep learning} and machine learning models currently learn to predict scientific classes from images \rev{(or similarly high dimensional data)}. Learning to encode these classes can be quite challenging as the concepts and definitions represented by these classes can be both abstract and contentious. This may be partially addressed by training models to encode a semantic taxonomy instead, as models would learn the features instead of abstracted classes. Derived \rev{semantic} taxonomies presents more clearly defined concepts and may encourage models to learn a more robust feature space. This could improve the effectiveness of the encoded features of a model for various other tasks. Even in a simple use case, the more robust feature space may allow models to be more generalisable and less brittle, i.e. transferable or fine-tunable to differing data sets and tasks, which would be of immediate benefit to a number of applications.

The \textit{anglocentric} nature of this work was alluded to previously. Although the language improvements, may benefit many populations, it does still marginalise those who do not speak English natively. The RGZ EMU team is considering a number of strategies to mitigate the effect of anglocentric labelling, including translation into multiple languages. However, we recognise that there are broader complex issues around use of language in science and recommend this as a topic of discussion in future work.

Finally, the \textit{ethics} of deriving terms from an unstructured data set include careful consideration of the potential presence and impact of malicious agents. Caution is therefore advised when applying this process to other fields. \rev{In this work,} the data set used was small enough that each annotation was inspected individually. 

\section{Conclusions}
\label{sec:conclusions}
In this work, we derive a flexible English taxonomy for radio astronomy and the respective morphological tagging. 
The proposed taxonomy \rev{of 22 semantic tags} is 
\begin{itemize}
    \item the product of experiments collecting expert classifications and plain English annotations on radio source morphologies using selected cutouts from the EMU pilot survey,
    \item reduced to a set of 10 terms to maximise its effectiveness within citizen science projects, starting with RGZ EMU,
    \item derived analytically through a novel method with minimal clarifying intervention.
\end{itemize}
\rev{We demonstrate the first effective use cases of the newly derived semantic morphology taxonomy. We show that using the tags we can recover 
\begin{itemize}
    \item known scientific morphologies, and
    \item rare sources with abnormal morphologies.
\end{itemize}}
The method which was developed, detailed, and applied in this work is domain agnostic. The method
\begin{itemize}
    \item provides a framework through which plain English annotations of complex ideas can return a ranked taxonomy on a given subject,
    \item can be applied to any scenarios where language is a barrier to future research,
    \item can increase the accessibility of complex scientific concepts by distilling concepts into simpler English for the public, collaborators, and citizen scientists.
\end{itemize}

The potential scientific impacts, applications, and communication benefits of this method and taxonomy
are discussed at length in Section~\ref{sec:discussion}.

\section*{Acknowledgements}
MB, MW, ELA, AMS, and IVS gratefully acknowledge support from the UK Alan Turing Institute under grant reference EP/V030302/1.
HT gratefully acknowledges the support from the Shuimu Tsinghua Scholar Program of Tsinghua University. EV acknowledges support by the Carl Zeiss Stiftung with the project code KODAR. ELA additionally gratefully acknowledges support from the UK Science \& Technology Facilities Council (STFC) under grant reference ST/P000649/1. AD acknowledges support by the BMBF Verbundforschung under the grant 05A20STA. DL acknowledges support from the Natural Sciences and Engineering Research Council of Canada.

This work has made use of data from the European Space Agency (ESA) mission
{\it Gaia} (\url{https://www.cosmos.esa.int/gaia}), processed by the {\it Gaia}
Data Processing and Analysis Consortium (DPAC,
\url{https://www.cosmos.esa.int/web/gaia/dpac/consortium}). Funding for the DPAC
has been provided by national institutions, in particular the institutions
participating in the {\it Gaia} Multilateral Agreement.

This research has made use of "Aladin sky atlas" developed at CDS, Strasbourg Observatory, France.

The Australian SKA Pathfinder is part of the Australia Telescope National Facility which is managed by CSIRO. Operation of ASKAP is funded by the Australian Government with support from the National Collaborative Research Infrastructure Strategy. ASKAP uses the resources of the Pawsey Supercomputing Centre. Establishment of ASKAP, the Murchison Radio-astronomy Observatory and the Pawsey Supercomputing Centre are initiatives of the Australian Government, with support from the Government of Western Australia and the Science and Industry Endowment Fund. We acknowledge the Wajarri Yamatji as the traditional owners of the Observatory site.

This project used public archival data from the Dark Energy Survey (DES). Funding for the DES Projects has been provided by the U.S. Department of Energy, the U.S. National Science Foundation, the Ministry of Science and Education of Spain, the Science and Technology Facilities Council of the United Kingdom, the Higher Education Funding Council for England, the National Center for Supercomputing Applications at the University of Illinois at Urbana-Champaign, the Kavli Institute of Cosmological Physics at the University of Chicago, the Center for Cosmology and Astro-Particle Physics at the Ohio State University, the Mitchell Institute for Fundamental Physics and Astronomy at Texas A\&M University, Financiadora de Estudos e Projetos, Funda{\c c}{\~a}o Carlos Chagas Filho de Amparo {\`a} Pesquisa do Estado do Rio de Janeiro, Conselho Nacional de Desenvolvimento Cient{\'i}fico e Tecnol{\'o}gico and the Minist{\'e}rio da Ci{\^e}ncia, Tecnologia e Inova{\c c}{\~a}o, the Deutsche Forschungsgemeinschaft, and the Collaborating Institutions in the Dark Energy Survey.

The Collaborating Institutions are Argonne National Laboratory, the University of California at Santa Cruz, the University of Cambridge, Centro de Investigaciones Energ{\'e}ticas, Medioambientales y Tecnol{\'o}gicas-Madrid, the University of Chicago, University College London, the DES-Brazil Consortium, the University of Edinburgh, the Eidgen{\"o}ssische Technische Hochschule (ETH) Z{\"u}rich,  Fermi National Accelerator Laboratory, the University of Illinois at Urbana-Champaign, the Institut de Ci{\`e}ncies de l'Espai (IEEC/CSIC), the Institut de F{\'i}sica d'Altes Energies, Lawrence Berkeley National Laboratory, the Ludwig-Maximilians Universit{\"a}t M{\"u}nchen and the associated Excellence Cluster Universe, the University of Michigan, the National Optical Astronomy Observatory, the University of Nottingham, The Ohio State University, the OzDES Membership Consortium, the University of Pennsylvania, the University of Portsmouth, SLAC National Accelerator Laboratory, Stanford University, the University of Sussex, and Texas A\&M University.

Based in part on observations at Cerro Tololo Inter-American Observatory, National Optical Astronomy Observatory, which is operated by the Association of Universities for Research in Astronomy (AURA) under a cooperative agreement with the National Science Foundation.

We thank the citizen scientists for their time and effort. We especially thank Victor Linares Sisifolibre for their substantial contributions.

\section*{Data Availability}

All the data used in this work is publicly available. 
The cutouts presented to users for annotation, anonymised annotations, and anonymised expert classifications are available at \url{https://zenodo.org/record/7254123\#.Y3d5EdLP2xE}.
The code written for these projects are available under \url{https://github.com/mb010/Text2Tag}.



\bibliographystyle{mnras}
\bibliography{example} 

\begin{thebibliography}{}
\makeatletter
\relax
\def\mn@urlcharsother{\let\do\@makeother \do\$\do\&\do\#\do\^\do\_\do\%\do\~}
\def\mn@doi{\begingroup\mn@urlcharsother \@ifnextchar [ {\mn@doi@}
  {\mn@doi@[]}}
\def\mn@doi@[#1]#2{\def\@tempa{#1}\ifx\@tempa\@empty \href
  {http://dx.doi.org/#2} {doi:#2}\else \href {http://dx.doi.org/#2} {#1}\fi
  \endgroup}
\def\mn@eprint#1#2{\mn@eprint@#1:#2::\@nil}
\def\mn@eprint@arXiv#1{\href {http://arxiv.org/abs/#1} {{\tt arXiv:#1}}}
\def\mn@eprint@dblp#1{\href {http://dblp.uni-trier.de/rec/bibtex/#1.xml}
  {dblp:#1}}
\def\mn@eprint@#1:#2:#3:#4\@nil{\def\@tempa {#1}\def\@tempb {#2}\def\@tempc
  {#3}\ifx \@tempc \@empty \let \@tempc \@tempb \let \@tempb \@tempa \fi \ifx
  \@tempb \@empty \def\@tempb {arXiv}\fi \@ifundefined
  {mn@eprint@\@tempb}{\@tempb:\@tempc}{\expandafter \expandafter \csname
  mn@eprint@\@tempb\endcsname \expandafter{\@tempc}}}

\bibitem[\protect\citeauthoryear{{Abbott} et~al.,}{{Abbott}
  et~al.}{2018}]{DES2018}
{Abbott} T.~M.~C.,  et~al., 2018, \mn@doi [\apjs] {10.3847/1538-4365/aae9f0},
  \href {https://ui.adsabs.harvard.edu/abs/2018ApJS..239...18A} {239, 18}

\bibitem[\protect\citeauthoryear{Adajian}{Adajian}{2022}]{Thomas2022art-definition}
Adajian T.,  2022, in Zalta E.~N.,  ed., , The {Stanford} Encyclopedia of
  Philosophy, {S}pring 2022 edn, Metaphysics Research Lab, Stanford University

\bibitem[\protect\citeauthoryear{{Banfield} et~al.,}{{Banfield}
  et~al.}{2015}]{Banfield+2015}
{Banfield} J.~K.,  et~al., 2015, \mn@doi [\mnras] {10.1093/mnras/stv1688},
  \href {https://ui.adsabs.harvard.edu/abs/2015MNRAS.453.2326B} {453, 2326}

\bibitem[\protect\citeauthoryear{{Boch} \& {Fernique}}{{Boch} \&
  {Fernique}}{2014}]{AladinLite2014}
{Boch} T.,  {Fernique} P.,  2014, in {Manset} N.,  {Forshay} P.,  eds,
  Astronomical Society of the Pacific Conference Series Vol. 485, Astronomical
  Data Analysis Software and Systems XXIII. p.~277

\bibitem[\protect\citeauthoryear{{Bonaldi}, {Bonato}, {Galluzzi}, {Harrison},
  {Massardi}, {Kay}, {De Zotti}  \& {Brown}}{{Bonaldi}
  et~al.}{2019}]{Bonaldi+2019}
{Bonaldi} A.,  {Bonato} M.,  {Galluzzi} V.,  {Harrison} I.,  {Massardi} M.,
  {Kay} S.,  {De Zotti} G.,   {Brown} M.~L.,  2019, \mn@doi [\mnras]
  {10.1093/mnras/sty2603}, \href
  {https://ui.adsabs.harvard.edu/abs/2019MNRAS.482....2B} {482, 2}

\bibitem[\protect\citeauthoryear{{Bonnarel} et~al.,}{{Bonnarel}
  et~al.}{2000}]{Aladin2000}
{Bonnarel} F.,  et~al., 2000, \mn@doi [\aaps] {10.1051/aas:2000331}, \href
  {https://ui.adsabs.harvard.edu/abs/2000A&AS..143...33B} {143, 33}

\bibitem[\protect\citeauthoryear{{Bowles} et~al.,}{{Bowles}
  et~al.}{2022}]{Bowles2022_SemanticTags}
{Bowles} M.,  et~al., 2022, NeurIPS 2022: Machine Learning and the Physical
  Sciences Workshop, \href
  {https://ui.adsabs.harvard.edu/abs/2022arXiv221014760B} {p. arXiv:2210.14760}

\bibitem[\protect\citeauthoryear{Crameri}{Crameri}{2021}]{cramerifabio}
Crameri F.,  2021, Scientific colour maps, \mn@doi{10.5281/zenodo.5501399},
  \url {https://doi.org/10.5281/zenodo.5501399}

\bibitem[\protect\citeauthoryear{{Dewdney}, {Hall}, {Schilizzi}  \&
  {Lazio}}{{Dewdney} et~al.}{2009}]{Dewdney2009SKA}
{Dewdney} P.~E.,  {Hall} P.~J.,  {Schilizzi} R.~T.,   {Lazio} T.~J.~L.~W.,
  2009, \mn@doi [IEEE Proceedings] {10.1109/JPROC.2009.2021005}, \href
  {https://ui.adsabs.harvard.edu/abs/2009IEEEP..97.1482D} {97, 1482}

\bibitem[\protect\citeauthoryear{{Fanaroff} \& {Riley}}{{Fanaroff} \&
  {Riley}}{1974}]{FR1974MNRAS.167P..31F}
{Fanaroff} B.~L.,  {Riley} J.~M.,  1974, \mn@doi [\mnras]
  {10.1093/mnras/167.1.31P}, \href
  {https://ui.adsabs.harvard.edu/abs/1974MNRAS.167P..31F} {167, 31P}

\bibitem[\protect\citeauthoryear{{Gaia Collaboration} et~al.,}{{Gaia
  Collaboration} et~al.}{2016}]{Gaia2016}
{Gaia Collaboration} et~al., 2016, \mn@doi [\aap]
  {10.1051/0004-6361/201629272}, \href
  {https://ui.adsabs.harvard.edu/abs/2016A&A...595A...1G} {595, A1}

\bibitem[\protect\citeauthoryear{{Gaia Collaboration} et~al.,}{{Gaia
  Collaboration} et~al.}{2021}]{Gaia2021}
{Gaia Collaboration} et~al., 2021, \mn@doi [\aap]
  {10.1051/0004-6361/202039657}, \href
  {https://ui.adsabs.harvard.edu/abs/2021A&A...649A...1G} {649, A1}

\bibitem[\protect\citeauthoryear{Grezes et~al.,}{Grezes
  et~al.}{2021}]{Grezes2021astroBert}
Grezes F.,  et~al., 2021. ADASS XXXI (31st annual conference on Astronomical
  Data Analysis Software and Systems), \mn@doi{10.48550/ARXIV.2112.00590.
}, \url {https://arxiv.org/abs/2112.00590}

\bibitem[\protect\citeauthoryear{Hagberg, Schult  \& Swart}{Hagberg
  et~al.}{2008}]{SciPyProceedings_11}
Hagberg A.~A.,  Schult D.~A.,   Swart P.~J.,  2008, in Varoquaux G.,  Vaught
  T.,   Millman J.,  eds, Proceedings of the 7th Python in Science Conference.
  Pasadena, CA USA, pp 11 -- 15

\bibitem[\protect\citeauthoryear{{Hallinan}, {Ravi}  \& {Deep Synoptic Array
  Team}}{{Hallinan} et~al.}{2021}]{Hallinan2021DSA2000}
{Hallinan} G.,  {Ravi} V.,   {Deep Synoptic Array Team} 2021, in American
  Astronomical Society Meeting Abstracts. p. 316.05

\bibitem[\protect\citeauthoryear{Hardcastle \& Croston}{Hardcastle \&
  Croston}{2020}]{Hardcastle2020}
Hardcastle M.~J.,  Croston J.~H.,  2020, \mn@doi [New Astronomy Reviews]
  {10.1016/J.NEWAR.2020.101539}, 88, 101539

\bibitem[\protect\citeauthoryear{{Hewish}, {Bell}, {Pilkington}, {Scott}  \&
  {Collins}}{{Hewish} et~al.}{1968}]{Hewish1968}
{Hewish} A.,  {Bell} S.~J.,  {Pilkington} J.~D.~H.,  {Scott} P.~F.,   {Collins}
  R.~A.,  1968, \mn@doi [\nat] {10.1038/217709a0}, \href
  {https://ui.adsabs.harvard.edu/abs/1968Natur.217..709H} {217, 709}

\bibitem[\protect\citeauthoryear{Honnibal \& Montani}{Honnibal \&
  Montani}{2017}]{spacy2}
Honnibal M.,  Montani I.,  2017, {spaCy 2}: Natural language understanding with
  {B}loom embeddings, convolutional neural networks and incremental parsing, To
  appear

\bibitem[\protect\citeauthoryear{Hunter}{Hunter}{2007}]{Hunter:2007}
Hunter J.~D.,  2007, \mn@doi [Computing in Science \& Engineering]
  {10.1109/MCSE.2007.55}, 9, 90

\bibitem[\protect\citeauthoryear{{Johnston} et~al.,}{{Johnston}
  et~al.}{2008}]{Johnston2008ASKAP}
{Johnston} S.,  et~al., 2008, \mn@doi [Experimental Astronomy]
  {10.1007/s10686-008-9124-7}, \href
  {https://ui.adsabs.harvard.edu/abs/2008ExA....22..151J} {22, 151}

\bibitem[\protect\citeauthoryear{{Jonas} \& {MeerKAT Team}}{{Jonas} \& {MeerKAT
  Team}}{2016}]{Jonas2016MeerKAT}
{Jonas} J.,  {MeerKAT Team} 2016, in MeerKAT Science: On the Pathway to the
  SKA. p.~1

\bibitem[\protect\citeauthoryear{{Kapi{\'n}ska} et~al.,}{{Kapi{\'n}ska}
  et~al.}{2017}]{kapinska+2017}
{Kapi{\'n}ska} A.~D.,  et~al., 2017, \mn@doi [\aj] {10.3847/1538-3881/aa90b7},
  \href {https://ui.adsabs.harvard.edu/abs/2017AJ....154..253K} {154, 253}

\bibitem[\protect\citeauthoryear{Lasker, Sturch, McLean, Russell, Jenkner  \&
  Shara}{Lasker et~al.}{1990}]{lasker1990guide}
Lasker B.~M.,  Sturch C.~R.,  McLean B.~J.,  Russell J.~L.,  Jenkner H.,
  Shara M.~M.,  1990, The Astronomical Journal, 99, 2019

\bibitem[\protect\citeauthoryear{Lundberg \& Lee}{Lundberg \&
  Lee}{2017}]{Lundberg2017SHAP}
Lundberg S.~M.,  Lee S.-I.,  2017, in Guyon I.,  Luxburg U.~V.,  Bengio S.,
  Wallach H.,  Fergus R.,  Vishwanathan S.,   Garnett R.,  eds, , Advances in
  Neural Information Processing Systems 30.
Curran Associates, Inc., pp 4765--4774, \url
  {http://papers.nips.cc/paper/7062-a-unified-approach-to-interpreting-model-predictions.pdf}

\bibitem[\protect\citeauthoryear{Lundberg et~al.,}{Lundberg
  et~al.}{2020}]{lundberg2020local2global}
Lundberg S.~M.,  et~al., 2020, Nature Machine Intelligence, 2, 2522

\bibitem[\protect\citeauthoryear{Lupton, Blanton, Fekete, Hogg, O’Mullane,
  Szalay  \& Wherry}{Lupton et~al.}{2004}]{lupton2004preparing}
Lupton R.,  Blanton M.~R.,  Fekete G.,  Hogg D.~W.,  O’Mullane W.,  Szalay
  A.,   Wherry N.,  2004, Publications of the Astronomical Society of the
  Pacific, 116, 133

\bibitem[\protect\citeauthoryear{Menze, Kelm, Masuch, Himmelreich, Bachert,
  Petrich  \& Hamprecht}{Menze et~al.}{2009}]{Menze2009}
Menze B.~H.,  Kelm B.~M.,  Masuch R.,  Himmelreich U.,  Bachert P.,  Petrich
  W.,   Hamprecht F.~A.,  2009, \mn@doi [BMC Bioinformatics]
  {10.1186/1471-2105-10-213/TABLES/4}, 10, 1

\bibitem[\protect\citeauthoryear{Mishra \& Kumar}{Mishra \&
  Kumar}{2020}]{mishra2020natural}
Mishra B.~K.,  Kumar R.,  2020, Natural Language Processing in Artificial
  Intelligence.
CRC Press

\bibitem[\protect\citeauthoryear{{Morabito} et~al.,}{{Morabito}
  et~al.}{2022}]{Morabito2022LOFAR_VLBI}
{Morabito} L.~K.,  et~al., 2022, \mn@doi [\aap] {10.1051/0004-6361/202140649},
  \href {https://ui.adsabs.harvard.edu/abs/2022A&A...658A...1M} {658, A1}

\bibitem[\protect\citeauthoryear{{Murphy} \& {ngVLA Science Advisory
  Council}}{{Murphy} \& {ngVLA Science Advisory Council}}{2020}]{ngVLA2020AAS}
{Murphy} E.,  {ngVLA Science Advisory Council} 2020, in American Astronomical
  Society Meeting Abstracts \#235. p. 364.01

\bibitem[\protect\citeauthoryear{Natal, Ávila, Tsukahara, Pinheiro  \&
  Maciel}{Natal et~al.}{2021}]{Jordao2021Entropy}
Natal J.,  Ávila I.,  Tsukahara V.~B.,  Pinheiro M.,   Maciel C.~D.,  2021,
  \mn@doi [Entropy] {10.3390/e23101340}, 23

\bibitem[\protect\citeauthoryear{{Norris} et~al.,}{{Norris}
  et~al.}{2011}]{Norris+2011}
{Norris} R.~P.,  et~al., 2011, \mn@doi [\pasa] {10.1071/AS11021}, \href
  {https://ui.adsabs.harvard.edu/abs/2011PASA...28..215N} {28, 215}

\bibitem[\protect\citeauthoryear{{Norris} et~al.,}{{Norris}
  et~al.}{2021a}]{Norris+2021a}
{Norris} R.~P.,  et~al., 2021a, \mn@doi [\pasa] {10.1017/pasa.2020.52}, \href
  {https://ui.adsabs.harvard.edu/abs/2021PASA...38....3N} {38, e003}

\bibitem[\protect\citeauthoryear{{Norris} et~al.,}{{Norris}
  et~al.}{2021b}]{Norris+2021b}
{Norris} R.~P.,  et~al., 2021b, \mn@doi [\pasa] {10.1017/pasa.2021.42}, \href
  {https://ui.adsabs.harvard.edu/abs/2021PASA...38...46N} {38, e046}

\bibitem[\protect\citeauthoryear{Pedregosa et~al.,}{Pedregosa
  et~al.}{2011}]{scikit-learn}
Pedregosa F.,  et~al., 2011, Journal of Machine Learning Research, 12, 2825

\bibitem[\protect\citeauthoryear{Pennington, Socher  \& Manning}{Pennington
  et~al.}{2014}]{pennington2014glove}
Pennington J.,  Socher R.,   Manning C.~D.,  2014, in Empirical Methods in
  Natural Language Processing (EMNLP). pp 1532--1543, \url
  {http://www.aclweb.org/anthology/D14-1162}

\bibitem[\protect\citeauthoryear{Romm}{Romm}{1991}]{Romm1991Utopia}
Romm J.,  1991, The Sixteenth Century Journal, 22, 173

\bibitem[\protect\citeauthoryear{{Rudnick}}{{Rudnick}}{2021}]{Rudnick+2021}
{Rudnick} L.,  2021, \mn@doi [Galaxies] {10.3390/galaxies9040085}, \href
  {https://ui.adsabs.harvard.edu/abs/2021Galax...9...85R} {9, 85}

\bibitem[\protect\citeauthoryear{Schelling}{Schelling}{1994}]{Schelling1994modernPhilosophy}
Schelling F. W. J.~V.,  1994, On the History of Modern Philosophy.
Cambridge University Press

\bibitem[\protect\citeauthoryear{{Schoenmakers}, {de Bruyn}, {R{\"o}ttgering},
  {van der Laan}  \& {Kaiser}}{{Schoenmakers} et~al.}{2000}]{Schoenmakers+2000}
{Schoenmakers} A.~P.,  {de Bruyn} A.~G.,  {R{\"o}ttgering} H.~J.~A.,  {van der
  Laan} H.,   {Kaiser} C.~R.,  2000, \mn@doi [\mnras]
  {10.1046/j.1365-8711.2000.03430.x}, \href
  {https://ui.adsabs.harvard.edu/abs/2000MNRAS.315..371S} {315, 371}

\bibitem[\protect\citeauthoryear{{Skrutskie} et~al.,}{{Skrutskie}
  et~al.}{2006}]{2006AJ....131.1163S}
{Skrutskie} M.~F.,  et~al., 2006, \mn@doi [\aj] {10.1086/498708}, \href
  {https://ui.adsabs.harvard.edu/abs/2006AJ....131.1163S} {131, 1163}

\bibitem[\protect\citeauthoryear{Southworth}{Southworth}{1956}]{Southworth1957RadioHistory}
Southworth G.~C.,  1956, The Scientific Monthly, 82, 55

\bibitem[\protect\citeauthoryear{Thomas, Thronson, Buonomo  \& Barbier}{Thomas
  et~al.}{2022}]{Thomas2022MLforSciencePlanning}
Thomas B.,  Thronson H.,  Buonomo A.,   Barbier L.,  2022, \mn@doi [Research
  Notes of the AAS] {10.3847/2515-5172/AC4990}, 6, 11

\bibitem[\protect\citeauthoryear{{Tingay} et~al.,}{{Tingay}
  et~al.}{2013}]{Tingay2013MWA}
{Tingay} S.~J.,  et~al., 2013, \mn@doi [\pasa] {10.1017/pasa.2012.007}, \href
  {https://ui.adsabs.harvard.edu/abs/2013PASA...30....7T} {30, e007}

\bibitem[\protect\citeauthoryear{Vayansky \& Kumar}{Vayansky \&
  Kumar}{2020}]{Vayansky2020TopicModellingReview}
Vayansky I.,  Kumar S. A.~P.,  2020, \mn@doi [Information Systems]
  {10.1016/j.is.2020.101582}, 94, 101582

\bibitem[\protect\citeauthoryear{Wald, Longo  \& Dobell}{Wald
  et~al.}{2016}]{Wald2016}
Wald D.~M.,  Longo J.,   Dobell A.~R.,  2016, \mn@doi [Conservation Biology]
  {10.1111/COBI.12627}, 30, 562

\bibitem[\protect\citeauthoryear{{Walmsley} et~al.,}{{Walmsley}
  et~al.}{2022}]{WalmsleyDECALSgz2022}
{Walmsley} M.,  et~al., 2022, \mn@doi [\mnras] {10.1093/mnras/stab2093}, \href
  {https://ui.adsabs.harvard.edu/abs/2022MNRAS.509.3966W} {509, 3966}

\bibitem[\protect\citeauthoryear{Waskom}{Waskom}{2021}]{Waskom2021}
Waskom M.~L.,  2021, \mn@doi [Journal of Open Source Software]
  {10.21105/joss.03021}, 6, 3021

\bibitem[\protect\citeauthoryear{{W}es {M}c{K}inney}{{W}es
  {M}c{K}inney}{2010}]{mckinney-proc-scipy-2010}
{W}es {M}c{K}inney 2010, in {S}t\'efan van~der {W}alt {J}arrod {M}illman eds,
  {P}roceedings of the 9th {P}ython in {S}cience {C}onference. pp 56 -- 61,
  \mn@doi{10.25080/Majora-92bf1922-00a}

\bibitem[\protect\citeauthoryear{Wolff \& Holmes}{Wolff \&
  Holmes}{2010}]{Wolff2010LinguisticRelativity}
Wolff P.,  Holmes K.~J.,  2010, \mn@doi [Ltd. WIREs Cogn Sci]
  {10.1002/wcs.104}, 2, 253

\bibitem[\protect\citeauthoryear{{Wright} et~al.,}{{Wright}
  et~al.}{2010}]{Wright2010WISE}
{Wright} E.~L.,  et~al., 2010, \mn@doi [\aj] {10.1088/0004-6256/140/6/1868},
  \href {https://ui.adsabs.harvard.edu/abs/2010AJ....140.1868W} {140, 1868}

\bibitem[\protect\citeauthoryear{pandas~development team}{pandas~development
  team}{2020}]{reback2020pandas}
pandas~development team T.,  2020, pandas-dev/pandas: Pandas,
  \mn@doi{10.5281/zenodo.3509134}, \url
  {https://doi.org/10.5281/zenodo.3509134}

\bibitem[\protect\citeauthoryear{{van Haarlem} et~al.,}{{van Haarlem}
  et~al.}{2013}]{LOFAR2013}
{van Haarlem} M.~P.,  et~al., 2013, \mn@doi [\aap]
  {10.1051/0004-6361/201220873}, \href
  {https://ui.adsabs.harvard.edu/abs/2013A&A...556A...2V} {556, A2}

\makeatother
\end{thebibliography}




\appendix
\section{Tag Definitions}\label{appendix:definitions}

The below are the tags which will be used in RGZ EMU, as discussed in Section~\ref{subsec:Tags for citizen science}. Here we present the same definitions of the tags as we are planning on providing to the citizen scientists:

\begin{itemize}
    \item \textit{Amorphous}: having no clearly defined shape or form.
    \item \textit{Bent}: Curved or having an angle.
    \item \textit{Bridge}: A structure that connects from one side to another (not a jet; see below).
    \item \textit{Core}: a central part distinct from the enveloping part.
    \item \textit{Hourglass}: Shaped like an hourglass.
    \item \textit{Jet}: A narrow stream of material appearing to emanate from a celestial object.
    \item \textit{Lobe}: A roundish projecting part of something divided by a fissure / gap.
    \item \textit{Merger}: Multiple separate things which appear to be connected or connecting.
    \item \textit{Plume}: a long cloud of smoke or vapour resembling a feather as it spreads from its point of origin.
    \item \textit{Tail}: Resembling an animal's tail in its shape or position, typically extending downwards or outwards at the end of a thing.
\end{itemize}

The tags which are proposed for algorithmic assignment according to Table~\ref{tab:Assigning tags} are not defined here. They will likely be defined when respective algorithms are developed.

\section{Star Forming Galaxies without `Traces Host Galaxy' Tag}
\label{appendix:SFG no tag} 
\begin{figure*}
    \centering
    \includegraphics[width=0.9\textwidth]{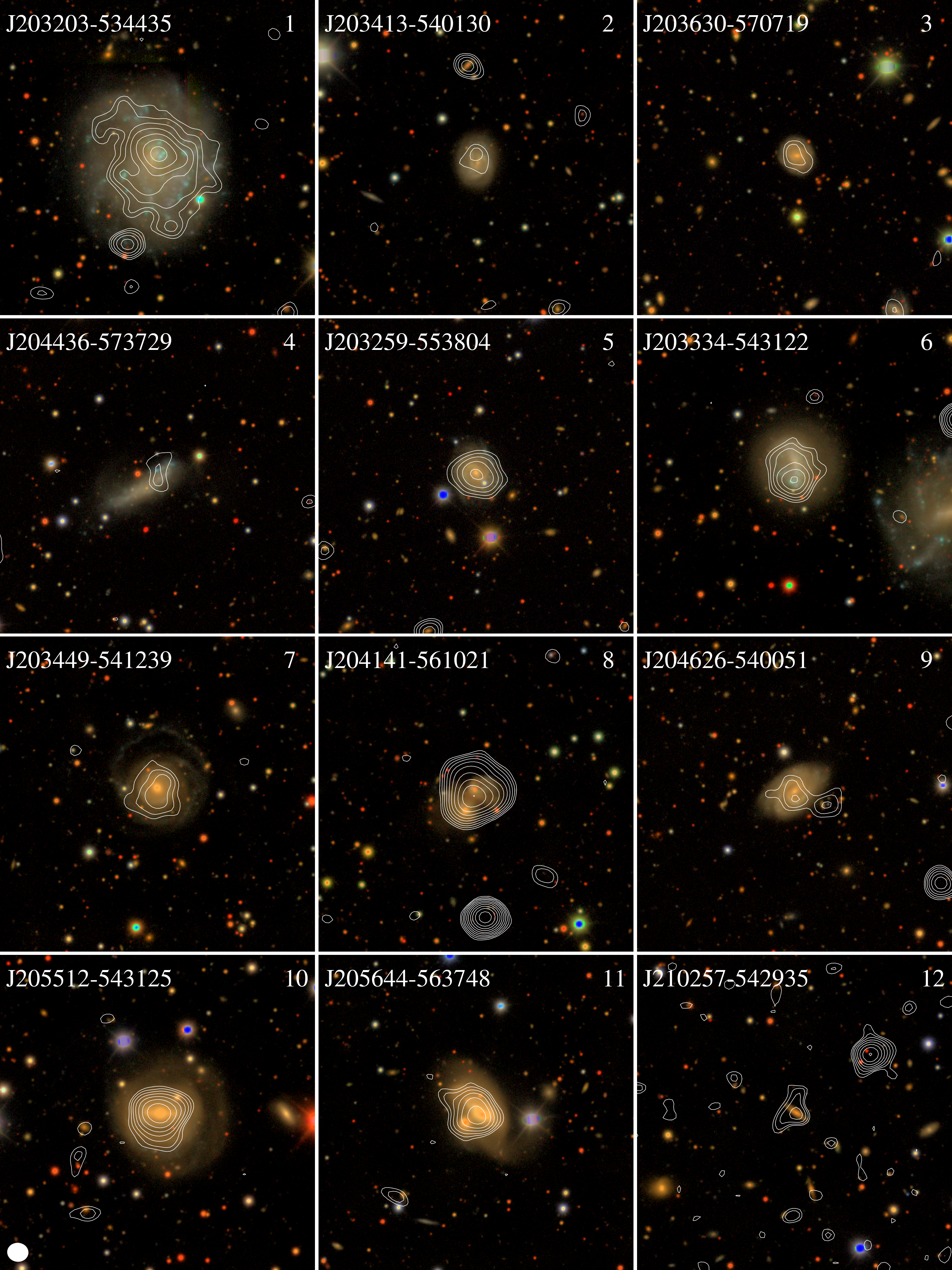}
    \caption{Composite images of SFG sources not captured by the `trace' tag, as discussed in Section~\ref{subsec:Detecting traditinoal Populations}. EMU contours following Figure~\ref{fig:cutout example} with optical DES RGB backgrounds as in Figure~\ref{fig:SFG_composite}. Cutout centre coordinates are presented on the image along side their respective source numbers associated with Table~\ref{tab:sfg no trace tag} and Table~\ref{tab:optical morphologies for sfg no tag}. The radio beam size for all panels is shown in the lower left of the figure; all cutouts are $3^\prime\times3^\prime$.}
    \label{fig:SFG no tag}
\end{figure*}

Figure~\ref{fig:SFG no tag} presents the sources which were classified (above 60\% agreement) as star forming galaxies, but were not tagged with `trace' in our synthetic catalogue, as described in Section~\ref{subsec:Detecting traditinoal Populations}. Table~\ref{tab:optical morphologies for sfg no tag} presents the optical morphologies of the sources listed in Table~\ref{tab:sfg no trace tag} produced by Walmsley et al. (in prep.), and made with Zoobot\footnote{\url{https://github.com/mwalmsley/zoobot}} \citep[initially described in][]{WalmsleyDECALSgz2022}. Here source numbers align with those presented in Table~\ref{tab:sfg no trace tag}.
\begin{table*}
\begin{tabular}{ccl}
\hline
No.                    & \begin{tabular}[c]{@{}c@{}}Coordinates\\ (J2000)\end{tabular} & GZ Morphology Predictions                                                                               \\ \hline
1  & 20h\,32m\,03s -53$^\circ44^\prime35^{\prime\prime}$ & featured (74\%; not edge-on), face-on (98\%), spiral (76\%), no bar (59\%), small bulge (59\%)                       \\
2  & 20h\,34m\,13s -54$^\circ01^\prime30^{\prime\prime}$ & featured (72\%), face-on (99\%; not edge-on), spiral (70\%), no bar (60\%), no bulge (69\%)                          \\
3  & 20h\,36m\,30s -57$^\circ07^\prime19^{\prime\prime}$ & featured (89\%), face-on (97\%; not edge-on), spiral (98\%), small bulge (85\%), tight arms (76\%)                   \\
4  & 20h\,44m\,36s -57$^\circ37^\prime29^{\prime\prime}$ & Not in catalogue. \\
5  & 20h\,32m\,59s -55$^\circ38^\prime04^{\prime\prime}$ & featured (87\%), face-on (98\%; not edge-on), spiral (98\%), no bar (77\%), small bulge (83\%), tight arms (84\%)    \\
6  & 20h\,33m\,34s -54$^\circ31^\prime22^{\prime\prime}$ & featured (70\%), face-on (95\%; not edge-on), no spiral (68\%)                                                       \\
7  & 20h\,34m\,49s -54$^\circ12^\prime39^{\prime\prime}$ & featured (82\%), face-on (97\%; not edge-on), spiral (85\%), no bar (76\%), moderate bulge (58\%), tight arms (63\%) \\
8  & 20h\,41m\,41s -56$^\circ10^\prime21^{\prime\prime}$ & featured (53\%), face-on (96\%; not edge-on), no spiral (77\%), no bar (84\%), merger (67\%)                         \\
9  & 20h\,46m\,26s -54$^\circ00^\prime51^{\prime\prime}$ & featured (83\%), face-on (98\%; not edge-on), spiral (85\%), small bulge (57\%)                                      \\
10 & 20h\,55m\,12s -54$^\circ31^\prime25^{\prime\prime}$ & Not in catalogue. \\
11 & 20h\,56m\,44s -56$^\circ37^\prime48^{\prime\prime}$ & featured (83\%), face-on (96\%; not edge-on), spiral (92\%), no bar (52\%), moderate bulge (59\%), tight arms (68\%) \\
12 & 21h\,02m\,57s -54$^\circ29^\prime35^{\prime\prime}$ & smooth (67\%), cigar shaped (88\%)                                                                        \\ \hline
\end{tabular}
\caption{Optical morphologies from Walmsley et al. (in prep.) of the SFG sources not captured by the `trace' tag, as discussed in Section~\ref{subsec:Detecting traditinoal Populations} and presented in Table~\ref{tab:sfg no trace tag}.}\label{tab:optical morphologies for sfg no tag}
\end{table*}

\section{Tag Rankings}
\label{appendix:Tag Rankings}

\begin{figure*}
    \centering
    \includegraphics[width=0.75\linewidth]{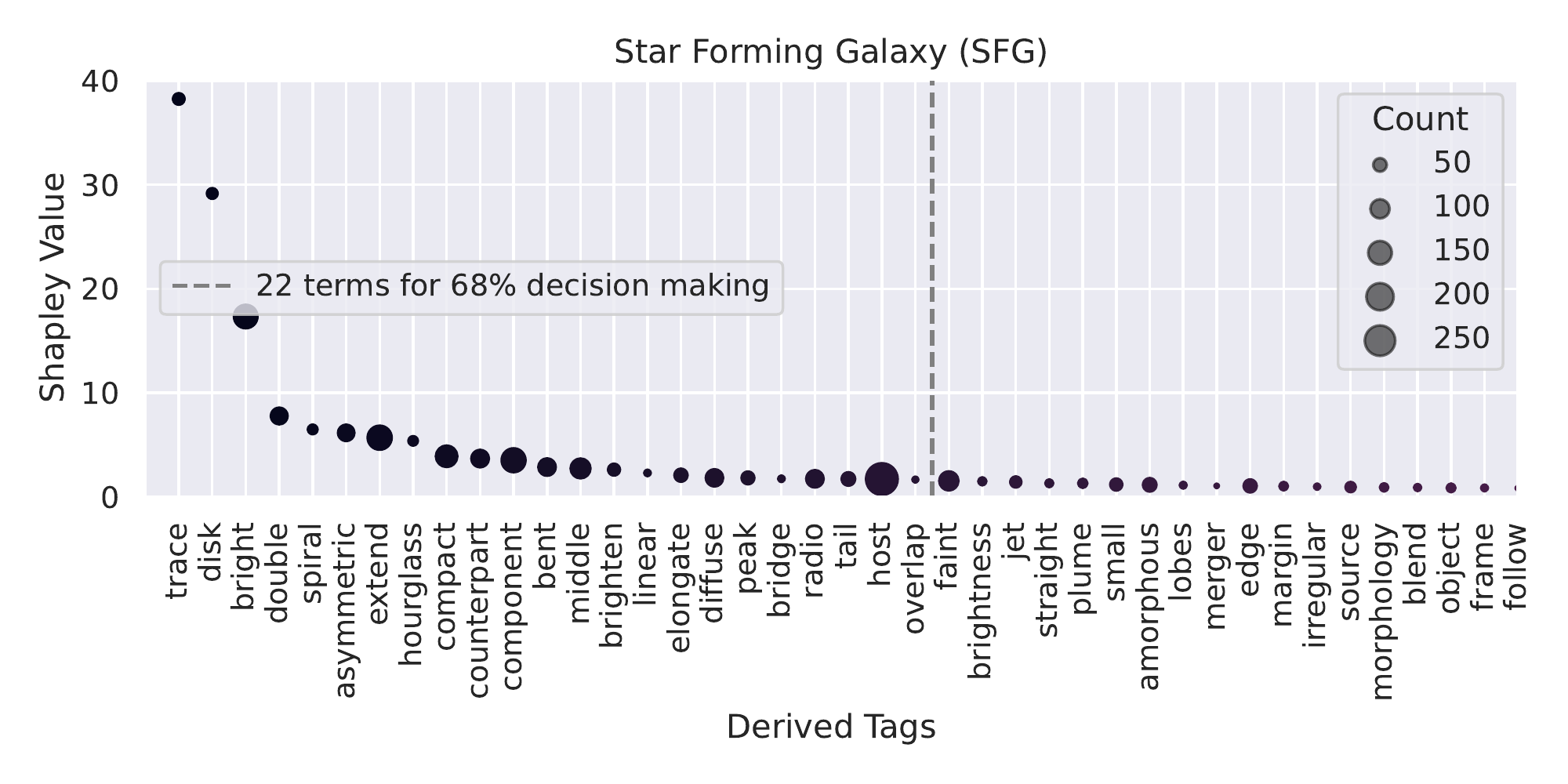}
    \caption{Class wise top Shapley value ranked tags for the Star-Forming Galaxy (SFG) science class.}
    \label{fig:SFG tag ranking}
\end{figure*}

To demonstrate the contributions of our tags to the classification of a given science class, we present Figure~\ref{fig:SFG tag ranking}, which shows the sorted most important tags for the Star-Forming Galaxy (SFG) class. Equivalent plots for each class are available in our public repository: \url{https://github.com/mb010/Text2Tag/data/AppendixB}.


\bsp	
\label{lastpage}
\end{document}